\documentclass{ws-ijmpc}

\pdfoutput=1

\usepackage{graphicx}
\usepackage{amsmath}

\newcommand{\rvec}{\boldsymbol{r}}
\newcommand{\evec}{\boldsymbol{e}}
\newcommand{\ivec}{\boldsymbol{i}}
\newcommand{\jvec}{\boldsymbol{j}}
\newcommand{\dvec}{\boldsymbol{d}}

\begin{document}

\title{Mountain trail formation and the active walker model}

\author{S. J. GILKS}
\address{Department of Physics, Loughborough University, LE11 3TU, UK}

\author{J. P. HAGUE}
\address{Department of Physics and Astronomy, The Open University, MK7 6AA, UK}

\date{9th January 2009}


\maketitle

\begin{abstract}

We extend the active walker model to address the formation of paths on
gradients, which have been observed to have a zigzag form. Our
extension includes a new rule which prohibits direct descent or ascent
on steep inclines, simulating aversion to falling. Further
augmentation of the model stops walkers from changing direction very
rapidly as that would likely lead to a fall. The extended model
predicts paths with qualitatively similar forms to the observed
trails, but only if the terms suppressing sudden direction changes are
included. The need to include terms into the model that stop rapid
direction change when simulating mountain trails indicates that a
similar rule should also be included in the standard active walker
model.


\end{abstract}

%
%

\ccode{PACS Nos.: 89.65.-s, 89.40.-a, 89.75.Kd}

\section{Introduction}

The dynamics of pedestrians and their interactions with the
environment have become a central theme in the study of social physics
\cite{helbing2001b}. An aspect of pedestrian dynamics that has
received considerable attention is the description of spontaneously
formed trail systems. Unexpected patterns can be found in trail
systems, highlighting the interplay between effective attractions and
itinerancy when walking from a starting point to a destination
\cite{helbing1997a}. The creation of powerful models of human trail
formation could help to improve planning and understanding of paths
and related phenomena
\cite{helbing1997a,helbing1997b,helbing2001a,kirchner2004a,goldstone2006a,goldstone2006b}. However,
we are not aware of any physics based studies of the formation of
mountain trail systems, where poor planning for human activity can
lead to significant environmental damage
\cite{bell1973a,coleman1981a}.

In principle, there are a huge number of possible routes to be
explored when choosing a path, and walkers could take any course
between a starting point and a destination. A first approximation to
the most probable path is a straight line between the initial and
destination points (unless there are obstacles in the way). However,
studies of trails using the active walker model have shown that the
detailed patterns of paths form from a counterpoint between the desire
to walk on well trodden paths, and the shortest route to be found by
traveling directly between the origin and destination
\cite{helbing1997a,helbing1997b}. Well-trodden paths are likely to be
favored by pedestrians because of reduced energy usage when compared,
for example, to walking through long grass. This preference may be
largely psychological, as internet based experiments in a virtual
environment have also shown that `walkers' tend to favor well-used
`paths' \cite{goldstone2006a,goldstone2006b}. The preference to walk on
regularly used paths leads to an effective interaction between
past and present walkers, indicating that there is interesting physics
involved in the formation of such trails.

\begin{figure*}
\includegraphics[height=63mm,angle=270]{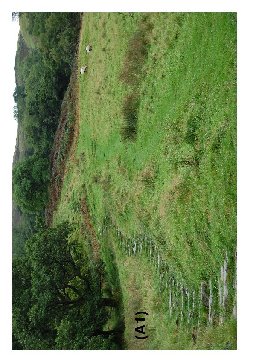}
\includegraphics[height=63mm,angle=270]{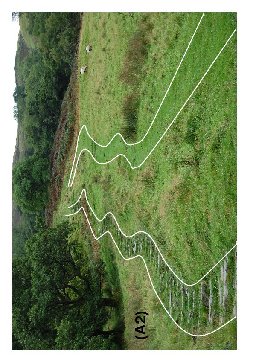}
\includegraphics[height=63mm,angle=270]{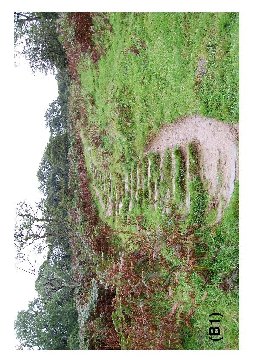}
\includegraphics[height=63mm,angle=270]{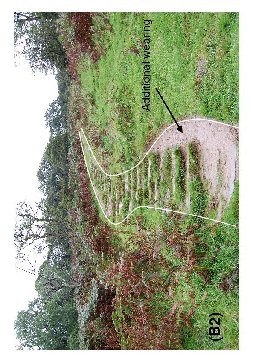}
\includegraphics[height=63mm,angle=270]{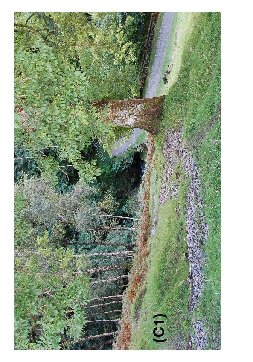}
\includegraphics[height=63mm,angle=270]{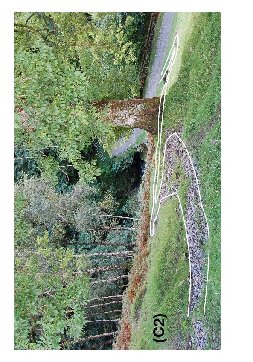}
\caption{(Color online) Spontaneously formed zig-zag paths on
Wansfell, near Ambleside, Lake District, Cumbria, UK. (A1, A2) The
left hand path has been augmented by humans since its formation. The
right hand path is spontaneously formed in the grass. Both paths have
a similar angle relative to the steepest direction and regularity of
direction change.). (B1, B2) Another zigzag path from further up the
trail. Recent wearing of the path can be seen. The top paths are on
inclines of around 1:8. The bottom path (figures C1, C2) is on a much
steeper incline of around 1:2. This path was found close to Low Sweden
Bridge near Ambleside, and breaks off from the main path that can be
seen curving around towards the back of the picture. White lines are
included on the right hand versions of the figures to highlight the
paths. \label{fig:lake}}
\end{figure*}

On inclines, there is a third influence.  Walkers may ascend slopes
diagonally if the gradient of the incline becomes too steep to ascend
directly. Walking at an angle to the line of fastest ascent has the
effect of decreasing the effective gradient of the incline, permitting
travel up steeper slopes. On descent, it may not be possible to walk
directly down a very steep slope without becoming unbalanced, again
leading walkers to take a diagonal path. Our aim in this article is to
determine if the walkers' desire to avoid steep gradients, in
combination with the rules of the active walker model, can be used to
simulate the zigzag paths that can be observed in mountainous regions.

The energetics of walking on an incline have been discussed by
Alexander in a simple model of bipedal locomotion
\cite{alexander2002a}. From energy considerations, walkers aim to
change the angle of ascent from the vertical if a hill reaches a steep
enough gradient. In Ref. \refcite{alexander2002a}, human walkers are
omniscient, and are able to assess the energetic outlay for an entire
route. In this way, they can assess if more energy would be expended
taking a shallower and longer route, or if a quick hike up a steep
route would be more favorable. We consider that it is unlikely that
the information about all possible routes is available for global
decisions to influence route and planning, and expect that decisions
are more likely to be local. Moreover, modeling has concluded that
humans choose well trodden paths in a more local manner
\cite{helbing2001a}. We therefore consider that insight into mountain
trail formation could be gained from active walker simulations of
inclined planes.

To demonstrate some examples of paths on inclines, we took photographs
in the Lake District in Cumbria, England, which can be seen in
Fig. \ref{fig:lake}. The upper panels Fig. \ref{fig:lake}(A1,A2) show
spontaneously formed zig-zag paths on Wansfell. Two paths can be seen
to the left and right hand sides of the picture, as highlighted in the
figure on the right (A2). The left hand path has been augmented with
rock since its formation, but still shows the characteristic
zig-zag. To remove doubt on the origin of the zig-zags (for example,
the stones might have been laid according to a plan) a second path can
be seen to the right of the picture. The right hand path is
spontaneously formed in the grass. Both have similar path angle and
regularity of direction change. An example from further up the trail
can be seen in panels B1 and B2. Here, there is additional wear to the
side of the trail, indicating that the trail is still evolving. The
paths are on inclines of around 1:8. The bottom path (C1 and C2) is on
an incline of around 1:2, and was observed near Low Sweden Bridge
which is close to the village of Ambleside. Trails of this type are
not unique to England, and such paths can be seen in other locations,
such as Smith Rock in the USA, where the bends in the paths are large
enough to appear on trail maps \cite{watts1992a}.

This article continues as follows. We review the rules of the active
walker model in section \ref{sec:activewalker}. In section
\ref{sec:gait} we briefly discuss aspects of the biomechanics of
walking on inclines. Extensions of the active walker model for
mountain trail systems are introduced in section
\ref{sec:mountainwalker}. Results of simulations are shown in section
\ref{sec:results}. Finally we summarize in section \ref{sec:summary}.

\section{Active walker model for human trails}
\label{sec:activewalker}

In order to investigate trail formations on mountains we use a
modified form of the active walker model, which was introduced in
Refs. \refcite{helbing1997a} and \refcite{helbing1997b}. In this
section, we review the rules of the unmodified active walker model
following the scheme introduced in Ref. \refcite{helbing1997a}. Since
the active walker model forms the basis of our extension to mountain
trails, it is our aim to ensure that all specific features of the
active walker model are clear, before introducing our extensions to
the model in section \ref{sec:mountainwalker}.

The aim of the active walker model is to describe the formation of
paths in soft ground. As pedestrians walk on soft surfaces such as
grass, the ground becomes worn and a path emerges. In the active
walker model, the wear on the soft surface is assumed to be
represented by a function, $G(\rvec,t)$, which represents the ground
condition at time $t$ and position $\rvec$. As is common in
statistical physics, $G(\rvec,t)$ is assumed to evolve according to a
first-order rate equation,
\begin{equation}
\frac{{\rm d}G({\rvec},t)}{{\rm d}t} = \frac{1}{T}\left[ G_{0} - G({\rvec},t) \right] + I\left[ 1 - \frac{G({\rvec},t)}{G_{\rm max}} \right] \sum_{\mu}\delta({\rvec} - {\rvec}_{\mu}).
\label{eqn:rate}
\end{equation} 
The rate of change of $G(\rvec,t)$ depends on weathering (the first
term on the right hand side of the equation) and wear by walkers (the
second term). The rate of weathering is governed by the parameter
$1/T$, where $T$ sets the time scale of path decay back to the
undisturbed ground condition $G_0$. The saturation value is denoted
$G_{\rm max}$. The parameter controlling the damage caused by a
footprint is denoted $I$. Walkers are located at positions
${\rvec}_{\mu}$, and $\delta({\rvec})$ is the Dirac delta function. In
the presence of more than one walker, $\mu$ is an index corresponding
to the walker being considered. Note that $G$ is a positive quantity
in this study.

The Dirac $\delta$-function found in equation \ref{eqn:rate} is not
especially convenient to use for numerics. Therefore, we suggest the
following small modification to equation \ref{eqn:rate},
\begin{equation}
\frac{{\rm d}G({\rvec},t)}{{\rm d}t} = \frac{1}{T}\left[ G_{0} -
G({\rvec},t) \right] + I\left[ 1 - \frac{G({\rvec},t)}{G_{\rm max}}
\right] \sum_{\mu}f({\rvec} - {\rvec}_{\mu}).
\label{eqn:ratenew}
\end{equation} 
where $f({\rvec} - {\rvec}_{\mu})$ is an arbitrary function that
specifies the shape of the damage caused by a footfall. $f({\rvec})$
is normalized to unity. We take $f$ to be a square with sides $l=10$cm
long, i.e. to have similar area to the base of a shoe. With this
choice of $f$, $l^2 G_{\rm max}/I$ is the approximate number of
footfalls that cause $G$ to reach the saturation value $G_{\rm
max}$. Since the feet of walkers have finite dimensions, we consider
equation \ref{eqn:ratenew} to be more physical than equation
\ref{eqn:rate}.

The positions of the walkers are updated according to how attracted
they are to any local paths, and the direction of the
destination. Each walker $\mu$ is influenced by a potential that
depends on their location relative to the paths. The attraction of
local paths is determined from,
\begin{equation}
V({\rvec}_{\mu},t) = \int {\rm d}^2{\rvec} \exp\left(-|{\rvec}-{\rvec}_{\mu}|/\sigma\right) G({\rvec},t).
\label{eqn:potential}
\end{equation}
The exponential function indicates how visible the path at ${\rvec}$
is to the pedestrian at location ${\rvec}_{\mu}$. The parameter,
$\sigma$ controls the distance over which walkers are motivated to
make detours to well formed trails. We note that the sign of the
potential $V$ is opposite to the usual convention, i.e. a positive $V$
indicates an attraction. There are no negative $V$, since $G$ is
always positive. This is in agreement with the sign convention used in
Ref. \refcite{helbing1997b}.

There are two competing influences on walkers in the active walker
model. The first is the desire to reach the destination in as direct a
manner as possible. In the conventional active walker model, each
walker with index $\mu$ aims for a destination point,
$\dvec_{\mu}$. In the absence of any paths, it is reasonable to expect
walkers to travel by the most direct route, i.e. along the unit
vector $(\dvec_{\mu}-\rvec_{\mu})/|\dvec_{\mu}-\rvec_{\mu}|$. The
attractive potential defined in equation \ref{eqn:potential} generates
an effective force in the same direction as the gradient of the
potential $\nabla_{\rvec_{\mu}}V(\rvec_{\mu},t)$. In the active walker
model, $\nabla_{\rvec_{\mu}}V(\rvec_{\mu},t)$ and
$(\dvec_{\mu}-\rvec_{\mu})/|\dvec_{\mu}-\rvec_{\mu}|$ are both used to
define the direction $\evec$ that walker $\mu$ moves along from
position $\rvec_{\mu}$ in the presence of a set of paths,
\begin{equation}
\evec(\rvec_{\mu},t) = \frac{(\dvec_{\mu} - \rvec_{\mu}(t))/|{\dvec}_{\mu} - {\rvec}_{\mu}(t)| + \nabla_{r_{\mu}}V_{\mu}(\rvec_{\mu},t)}{|(\dvec_{\mu} - \rvec_{\mu}(t))/|{\dvec}_{\mu} - {\rvec}_{\mu}(t)| + \nabla_{r_{\mu}}V_{\mu}({\rvec}_{\mu})|}
\label{eqn:unitvector}
\end{equation}
$\nabla_{r_{\mu}} V({\rvec}_{\mu},t)$ is the direction of favorable
ground. The symbol $\nabla_{r_{\mu}}$ represents the gradient of $V$
taken with respect to the position vector of the walker $\rvec_{\mu}$
(the subscript acts as a reminder that the gradient is not taken with
respect to $\rvec$). $\evec$ is a dimensionless unit vector.

The active walker model is completed with the equation of motion,
\begin{equation}
\frac{{\rm d}{\rvec}_{\mu}}{{\rm d}t} = v_{0\mu}{\evec}({\rvec}_{\mu},t)
\label{eqn:velocity}
\end{equation}
where $v_{0\mu}$ is the walking speed of walker $\mu$, assumed to
be constant for the entire journey of the walker. In the discretized
form of the equations, $v_{0\mu}$ is related to the length of a
stride.

The active walker model has been successful in describing some of the
unexpected features that have been observed in trail systems. The
application of the active walker model to the winding paths found on
steep inclines has not yet been considered. In the following section,
we discuss the biomechanics of walking on inclined planes with the aim
of developing additional rules to explain the wiggles observed in
trail systems on hills.

\section{The biomechanics of walking on inclines}
\label{sec:gait}

\begin{figure}
\includegraphics[width=75mm]{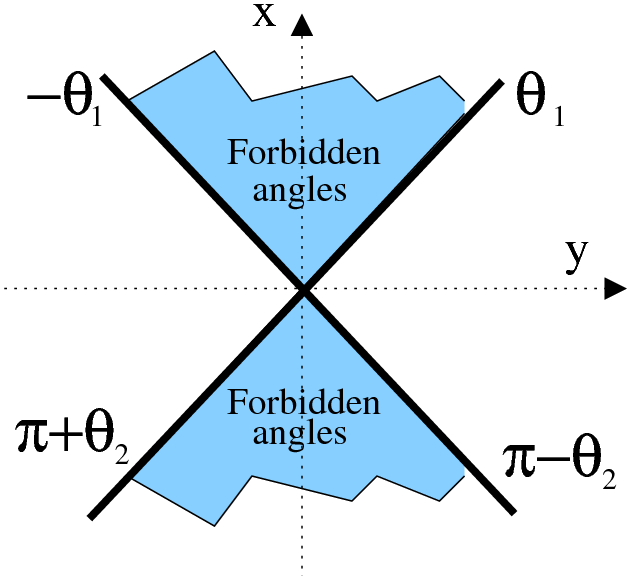}
\caption{Schematic of the region of discomfort for walking up or down
the mountain. If the active walker model suggests angles in the shaded
area $|\gamma|<\theta$, the angle will be mapped to either $\theta$ or
$-\theta$ depending on which side of the axis the optimally desired
angle lies. The mapping also depends on whether the walker is going up
or down the incline. We estimate that a forbidden angle of around
$10^o$ corresponds to a $1:4$ gradient. Note that the $x$-axis is
defined to be in the direction directly up the gradient, and the $y$
axis along the contour of no height change. Angles are defined from
the $x$-axis.}
\label{fig:modelsummary}
\end{figure}

In this section, we aim to give a brief summary of the physiological
factors that must be considered when modeling walkers on inclined
planes. Some insight into the manner of walking that is adopted on an
incline can be gained from attempts to develop bipedal robots that can
climb sloping surfaces \cite{zheng1990a}. If exactly the same method
is used to climb a slope as to walk on flat paths, the center of
gravity is positioned further back than usual, which may lead to a
fall \cite{zheng1990a} (a good introduction to the biomechanics of
walking on flat surfaces can be found in Ref.
\refcite{whittle2001a}). Humans (and robots developed to climb slopes)
compensate for the effects of the slope by leaning forward, which can
be achieved by flexing the ankle. This is satisfactory for very
shallow slopes. However, if a slope is too steep, it becomes
physiologically impossible to turn the ankle joint through the angle
necessary to hold the center of gravity over the feet (for example,
the person investigated in Ref. \refcite{leroux2002a} could not bend
his/her ankle by more than 24$^{o}$\footnote{In biomechanics, the
limit to bending of a joint is known as the maximum angular
excursion}). Thus good foot contact cannot be maintained with the ground
when walking directly up very steep inclines.

Experiments have shown that the requirements on joints increase
dramatically on moderate slopes \cite{leroux2002a,mcintosh2006a}:
walking on a 10$^{o}$ incline requires hip flexibility of 60$^{o}$,
compared with 30$^{o}$ on the flat. Demands on ankle flexibility
increase in a similar way. Joints in the leg such as the ankle, knee
and hip are also subjected to significantly increased forces
\cite{mcintosh2006a}. As determined in Ref. \refcite{leroux2002a}, the
ankle becomes fully bent for more of the walking cycle as the gradient
increases from $0^{o}$ to $10^{o}$. The physiological constraints on
the angles through which joints can bend indicate that hills
eventually become too steep to walk up directly. To compensate for the
limits of joint flexibility, the walker can choose to change the angle
of ascent, to avoid walking directly uphill. This leads to a smaller
effective gradient. The inability to walk directly uphill can be
thought of as an effective potential barrier against certain walking
directions.

There is also a risk of falling on descent. This is probably greater
than the chance of falling on ascent, since parts of the walking
cycle require the center of gravity to be moved forward
\cite{whittle2001a} and on descent, the body is already unstable to
falling forwards. Again, it is possible to compensate against falling
by reducing the angle of descent. As before, we expect this compensation
to lead to an effective potential barrier for angles that are likely
to cause a fall (most walkers will have learned to avoid
tumbling). When simulating mountain trails it is important to include
this behavior.

There is an additional feature of walking that should be relevant
regardless of whether the walker is on an incline or level
ground. Given the constraints of anatomy, sudden changes of direction
between consecutive footsteps are likely to lead to discomfort,
instability, falling or significant reduction in speed. Rapid changes
of direction correspond to significant changes in the momentum of the
walker and therefore large forces are required to maintain
stability. These forces are applied at ground level, so can lead to a
significant torque on the walker and toppling. We therefore expect
walkers to have a preference to make consecutive footsteps in similar
directions. We now describe the inclusion of incline effects and the
inability to make sudden changes of direction into the active walker
model.

\section{A model of mountain walkers}
\label{sec:mountainwalker}

\begin{figure}
\includegraphics[height=63mm,angle=270]{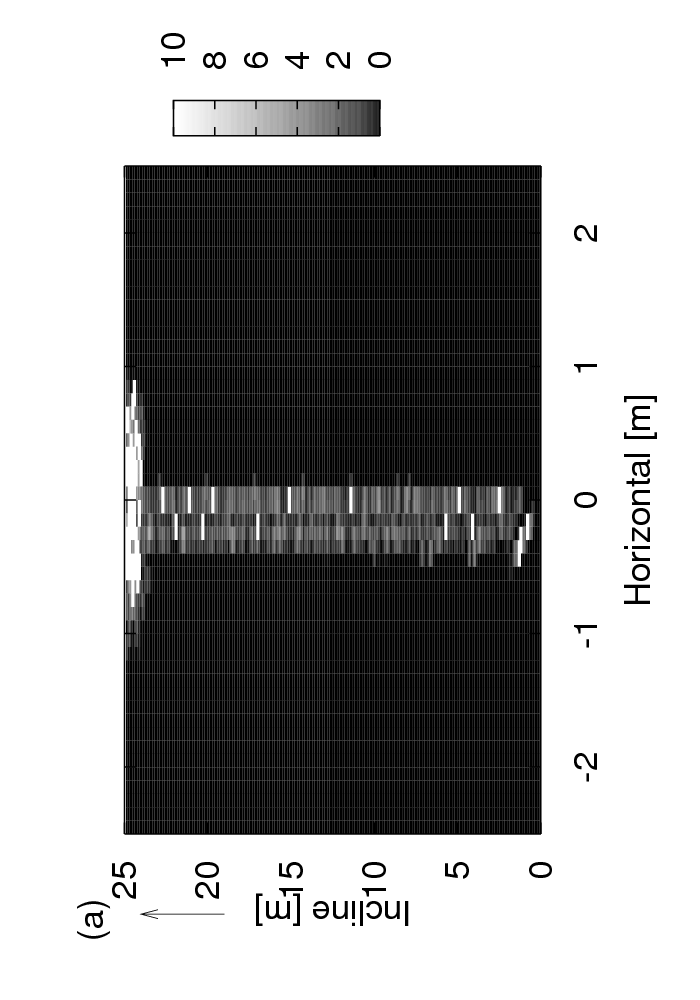}
\includegraphics[height=63mm,angle=270]{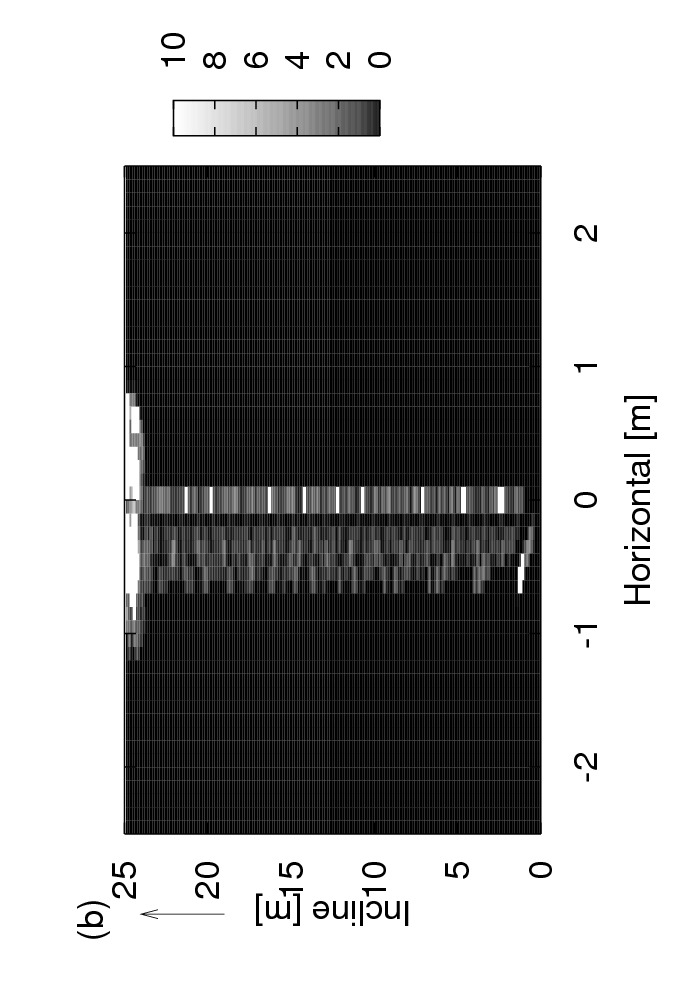}
\includegraphics[height=63mm,angle=270]{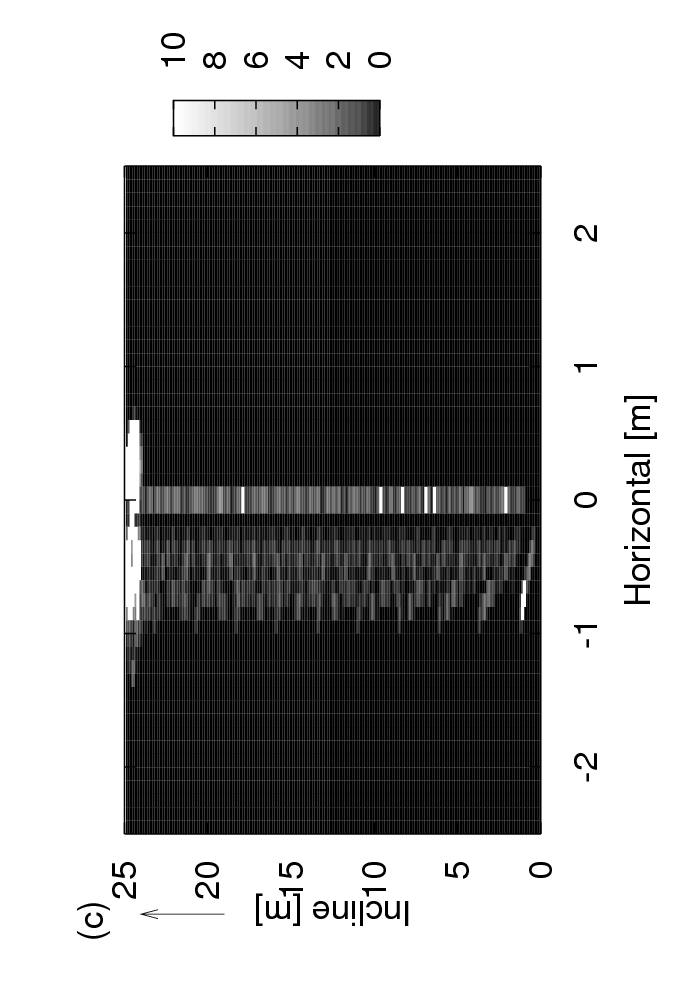}
\caption{Ground condition for (a) $\theta=15^o$ (b) $\theta=25^o$ and
(c) $\theta=35^o$. In all cases, the persistence of direction $\alpha
= 0$. If consecutive steps are not taken in similar directions, no
clear zig-zag paths form.}
\label{fig:mountain1}
\end{figure}

\begin{figure}
\includegraphics[height=63mm,angle=270]{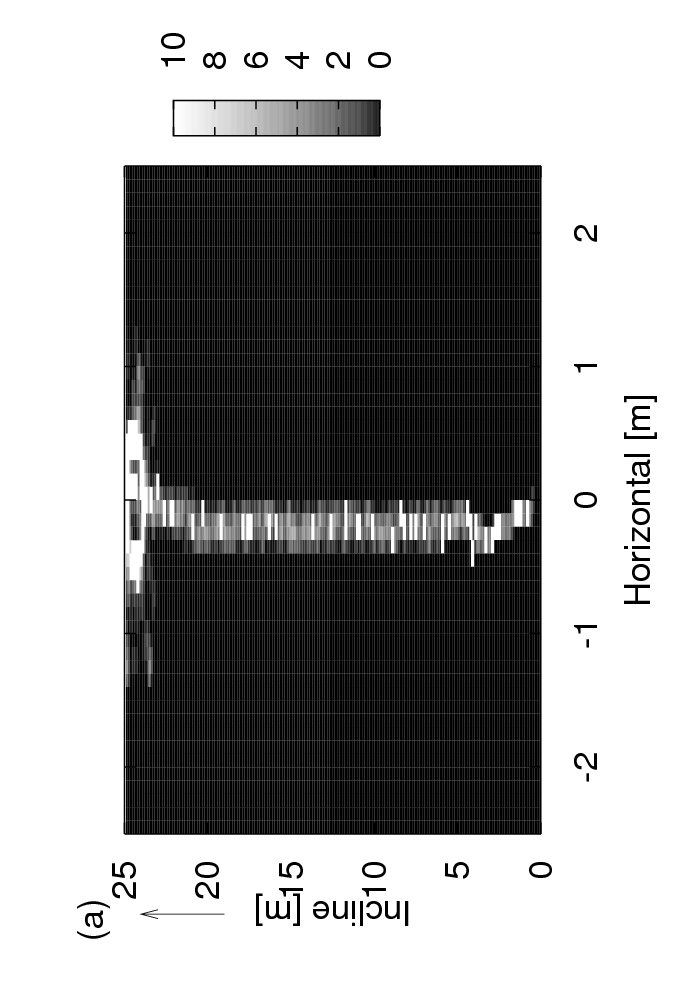}
\includegraphics[height=63mm,angle=270]{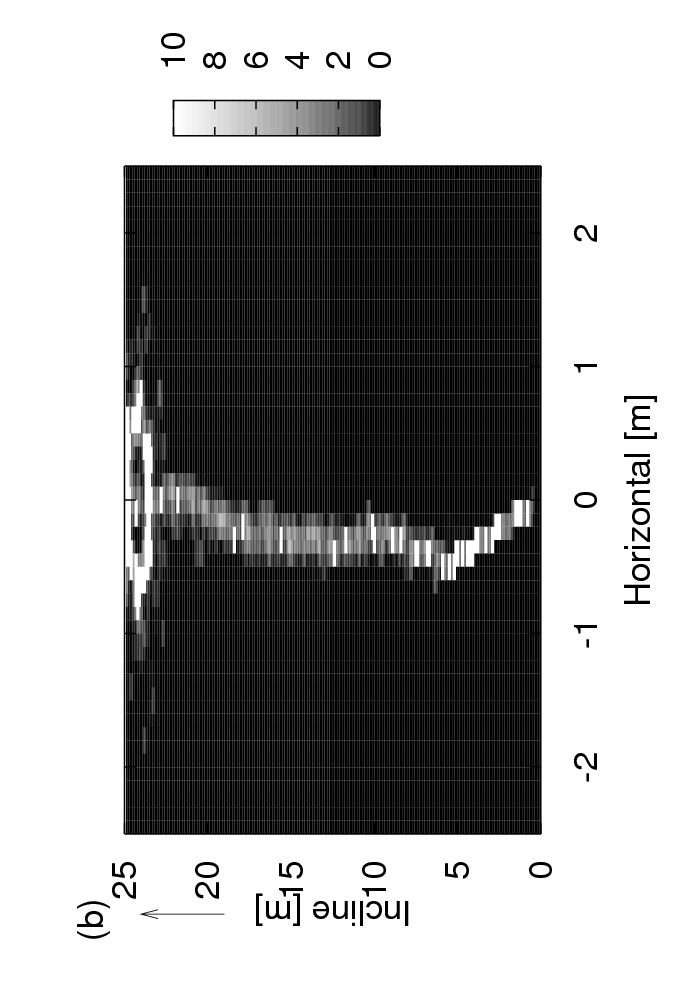}
\includegraphics[height=63mm,angle=270]{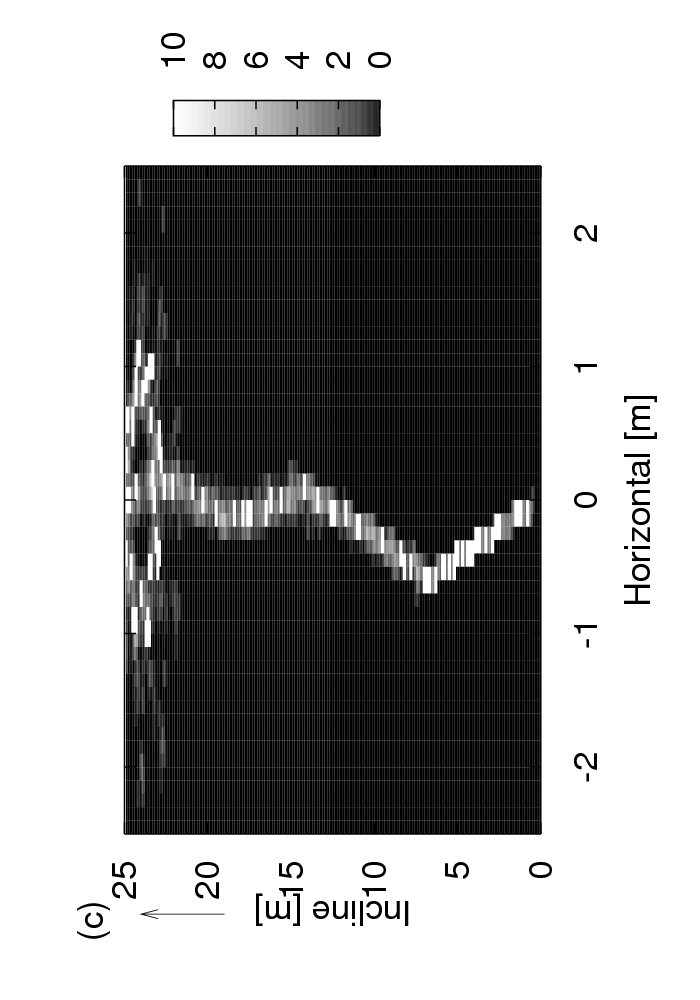}
\caption{Ground condition for simulations with minimum safe angle
$\theta=5^o$ and persistence of direction (a) $\alpha=0.4$ (b)
$\alpha=0.5$ and (c) $\alpha=0.6$. As the tendency to make consecutive
footfalls in the same direction increases, the size of the features
increases.}
\label{fig:mountainwalkerfive}
\end{figure}

\begin{figure}
\includegraphics[height=63mm,angle=270]{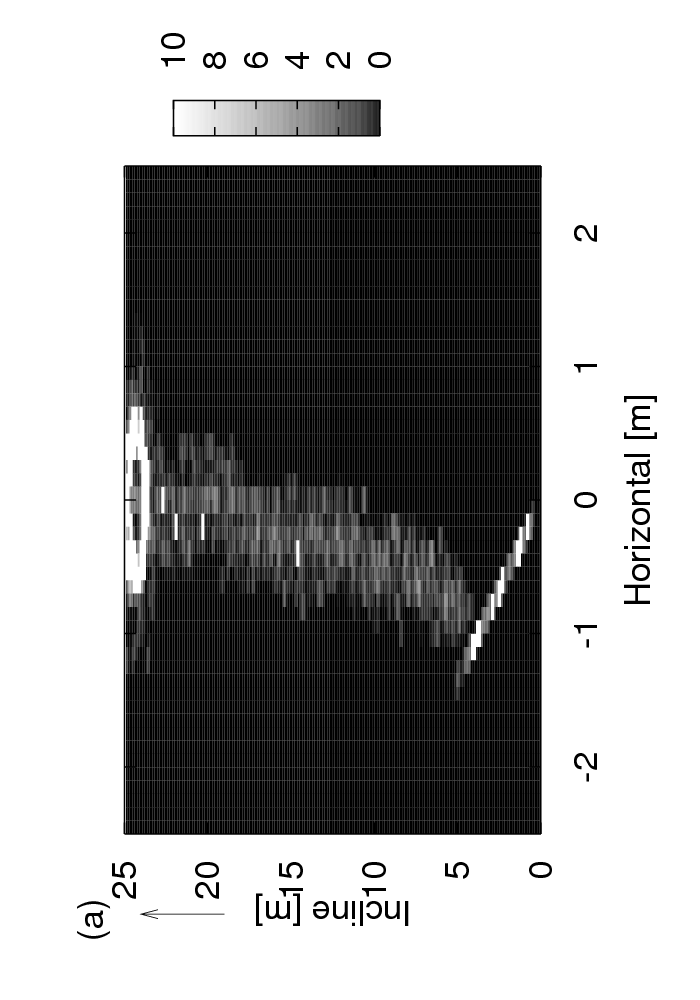}
\includegraphics[height=63mm,angle=270]{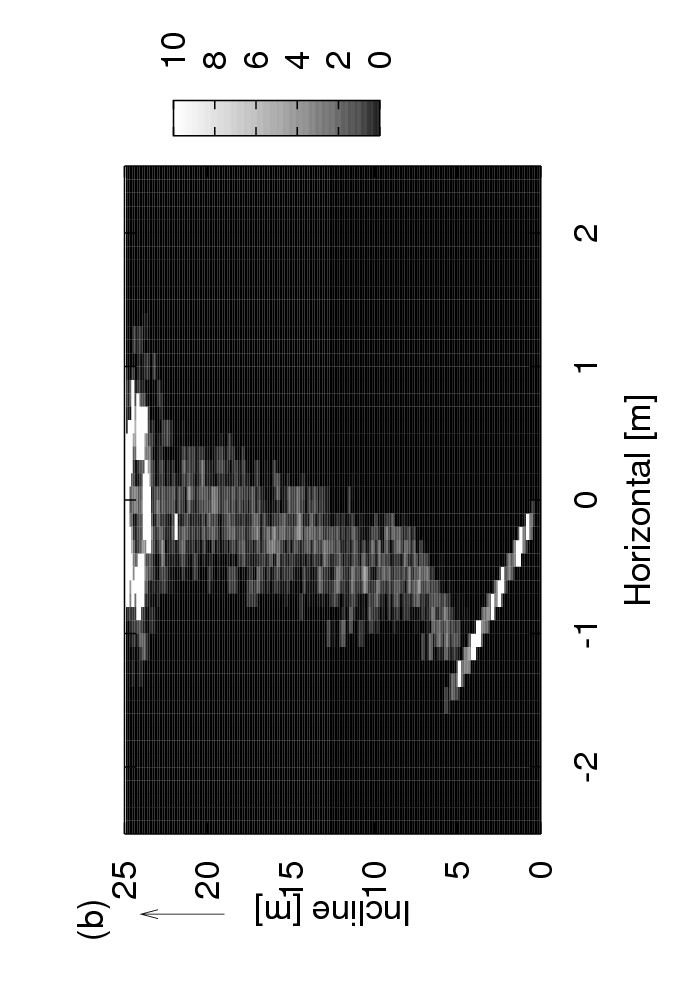}
\includegraphics[height=63mm,angle=270]{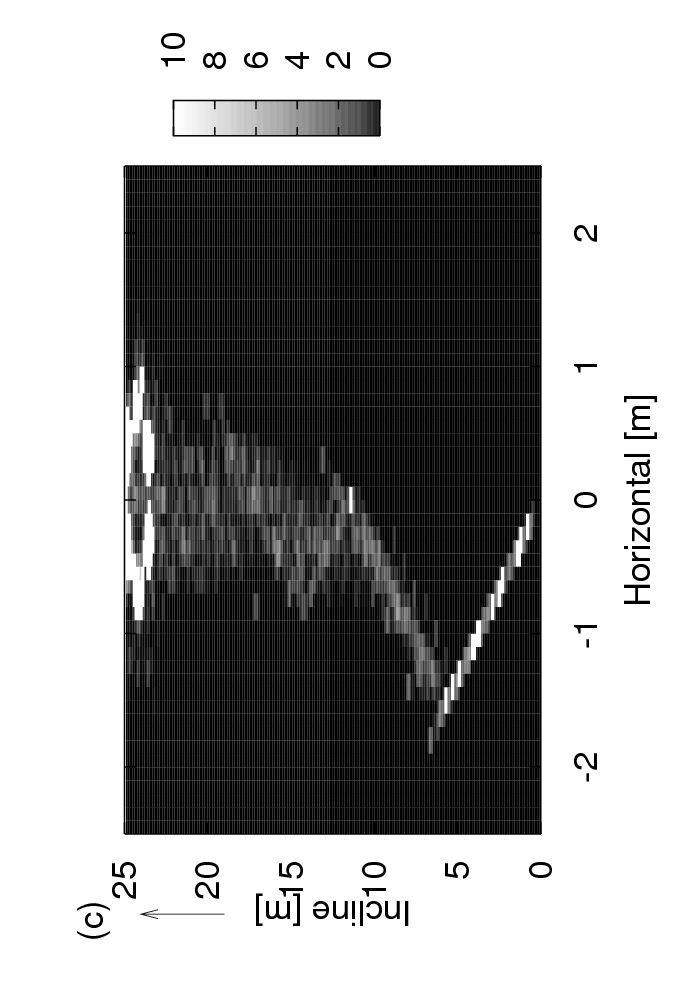}
\caption{Ground condition for simulations with $\theta = 15^o$ and (a)
$\alpha =0.4$ (b) $\alpha =0.45$ and (c) $\alpha =0.5$.}
\label{fig:mountainwalkerfifteen}
\end{figure}

\begin{figure}
\includegraphics[height=63mm,angle=270]{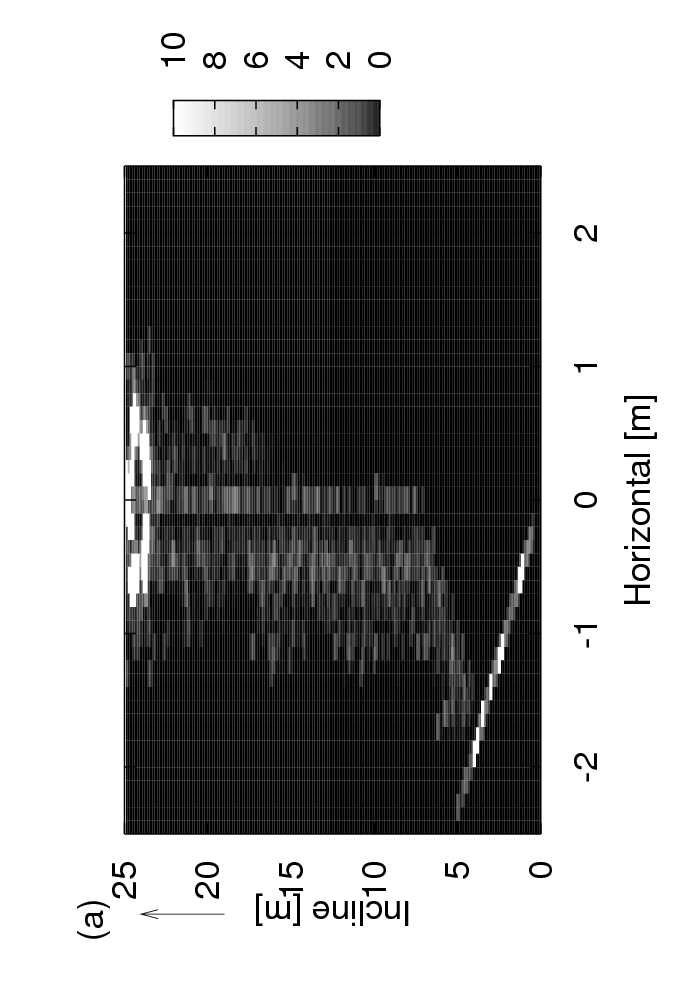}
\includegraphics[height=63mm,angle=270]{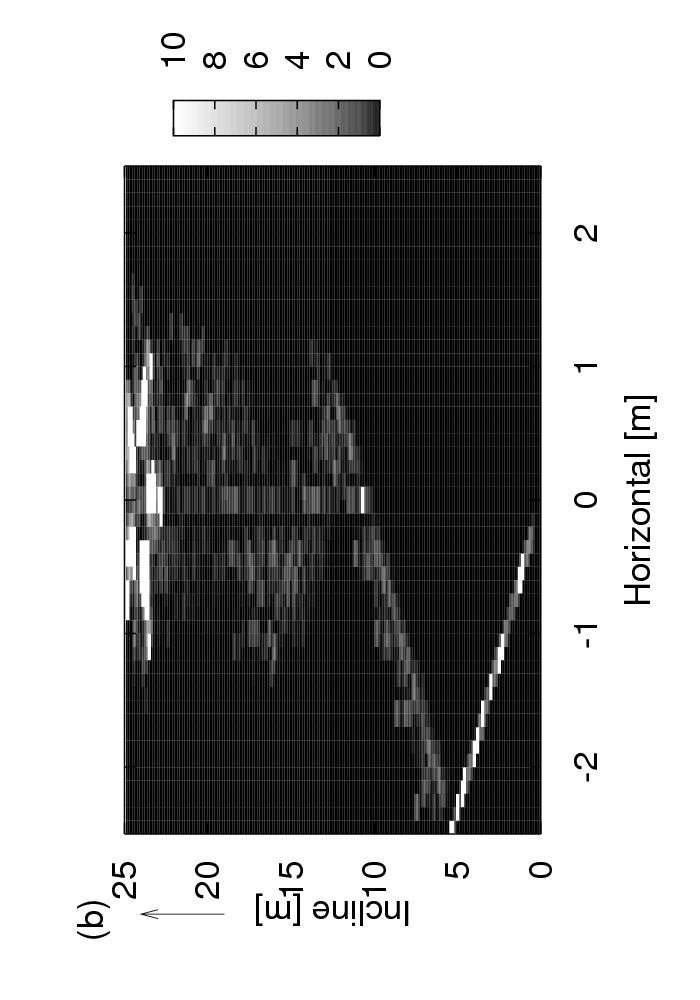}
\includegraphics[height=63mm,angle=270]{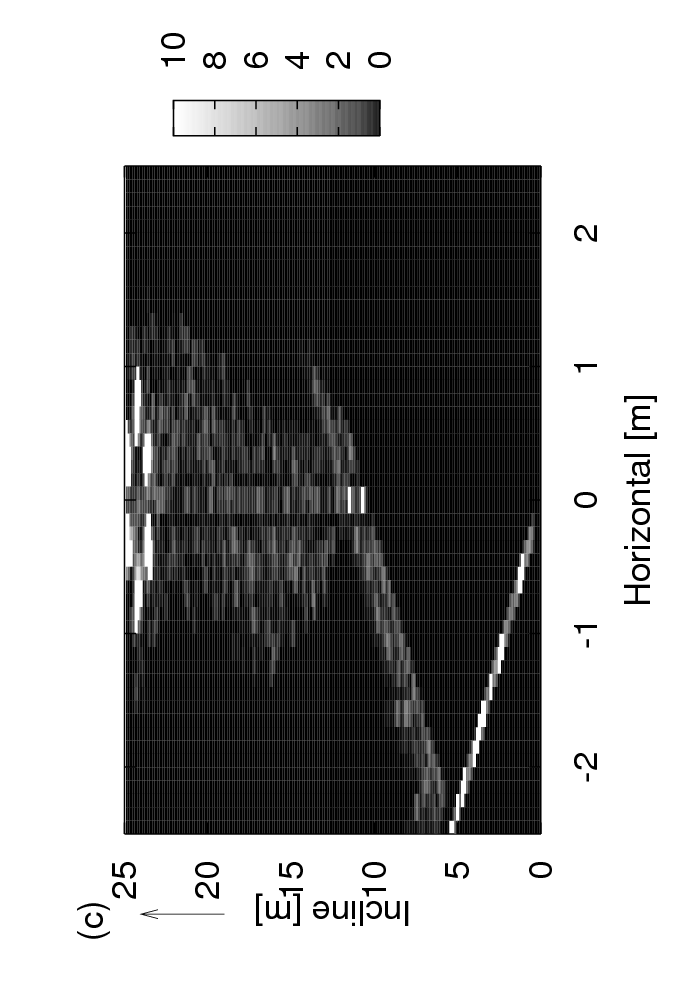}
\caption{Ground condition for simulations with $\theta=25^o$ and (a) $\alpha=0.4$ (b)
$\alpha=0.45$ and (c) $\alpha=0.5$.}
\label{fig:mountainwalkertwentyfive}
\end{figure}

\subsection{New rules for mountain walking}

It is our aim extend the active walker model to construct a ``mountain
walker model''. As we have discussed in section \ref{sec:gait}, our
model should take account of the inability of walkers on steep
inclines to walk directly up or down a slope if the gradient becomes
too great, and avoid sudden changes of direction that can cause
instability.


We suggest two ways of implementing an aversion to sudden changes in
direction:

\begin{description}
\item[Rule Ia] The new direction of the walker is taken as the weighted
average of the recent angle of motion $\phi$ and the angle of motion
that would be favored on a flat surface.
\begin{equation}
\gamma(t) = \alpha\phi + (1-\alpha)\beta(t),
\label{eqn:weighting}
\end{equation}

We call the parameter $\alpha$ the {\it persistence of
direction}. $\beta$ is the angle that is determined from the direction
vector $\evec = e_x \ivec + e_y \jvec$,
\begin{equation}
\beta(t) = \tan^{-1}(e_y(t) / e_x(t))
\label{eqn:inversetan}
\end{equation}
where $\ivec$ and $\jvec$ are unit vectors in the $x$ (uphill) and $y$
(horizontal) directions respectively. In the determination of $\beta$,
care is taken to ensure that the angle is in the quadrant consistent
with the signs of $e_x$ and $e_y$. We take care to avoid any problems
with branch cuts in equations \ref{eqn:weighting} and
\ref{eqn:inversetan}. Strictly, $\phi$ should relate to the direction
of the walker over the time period just before
\begin{equation}
\phi = \int_{t-\Delta}^{t} \beta(t') {\rm d}t'/\Delta
\label{eqn:averageangle}
\end{equation}
where $\Delta$ is the time period within which the direction of
previous footsteps is relevant to stability.
\end{description}

We also suggest an alternative to equation \ref{eqn:weighting}.

\begin{description}
\item[Rule Ib] Equation \ref{eqn:ratenew} is supplemented with the
average direction over recent time,
\begin{equation}
\evec(\rvec_{\mu},t) = \frac{(\dvec_{\mu} - \rvec_{\mu}(t))/|{\dvec}_{\mu} - {\rvec}_{\mu}(t)| + \nabla_{r_\mu}V_{\mu}(\rvec_{\mu})+\frac{\bar{\alpha}}{\Delta}\int_{t-\Delta}^{t}\evec(\rvec_{\mu},t'){\rm d} t'}{|(\dvec_{\mu} - \rvec_{\mu}(t))/|{\dvec}_{\mu} - {\rvec}_{\mu}(t)| + \nabla_{r_\mu}V_{\mu}({\rvec}_{\mu})+\frac{\bar{\alpha}}{\Delta}\int_{t-\Delta}^{t}\evec(\rvec_{\mu},t'){\rm d} t'|}.
\label{eqn:weightingnew}
\end{equation}
where again $\Delta$ is the time within which sudden change of
direction would lead to falling and $\bar{\alpha}$ is a parameter that
controls the influence of the previous footsteps.
\end{description}

We also modify the active walker model to incorporate the potential
barrier in angle space that represents the inability to achieve a safe
journey if the walker is traveling too close to the directly uphill
or downhill routes (i.e. a journey where the walker is unlikely to
fall over). This rule is summarized in figure \ref{fig:modelsummary}.

\begin{description}
\item[Rule IIa] (To be used with rule Ia) If the angle $\gamma$,
 chosen from equation \ref{eqn:weighting}, falls within a forbidden
 zone ($|\gamma|<\theta$), the angle along which the walker moves is
 mapped to the nearest minimum angle of safety (either $\theta$ or
 $-\theta$). We assume that the larger the incline of the slope, the
 larger the region of angles that are not permitted. This is
 equivalent to having an infinite effective potential barrier in
 certain directions. In the event that $\gamma=0$ our walker has a
 preference to move left. If there is a mapping, we reconstruct
 $\evec$ from $\gamma$ as $\evec=\cos(\gamma)\ivec+\sin(\gamma)\jvec$,
 noting that the sense of $\ivec$ is up the plane. The new $\evec$ is
 then used to update the position of the walker via
 eqn. \ref{eqn:velocity}.

\item[Rule IIb] (To be used with rule Ib) If the angle, $\beta = \tan^{-1}(e_y / e_x)$, lies
within a forbidden zone ($|\beta|<\theta$), then the vector $\evec$ is
mapped to the closest angle outside the forbidden region. If $\beta=0$
the walker moves left. Since $\beta=0$ on the first iteration, our
paths show a bias in that direction.
\end{description} 

\subsection{Discretization scheme}

We use the following simple first-order discretization procedure to
solve the active walker model and extensions. The time coordinate is
discretized into small intervals $\Delta t$, and the spatial
coordinates into intervals $\Delta x$ and $\Delta y$. Therefore, the
vector $\rvec$ can be written in the discrete form ${\rvec} =
(\eta\Delta x, \nu\Delta y)$ where $\eta$ and $\nu$ are
integers. Approximated using this scheme, equation \ref{eqn:ratenew}
becomes,
\begin{eqnarray}
G({\rvec},t_{n}) :=  G({\rvec},t_{n-1}) & + & \frac{\Delta t}{T}\left[G_0 - G({\rvec},t_{n-1})\right] \nonumber\\ 
&  & + \sum_{\mu}I\Delta t \left[ 1 - \frac{G({\rvec},t_{n-1})}{G_{\rm max}}\right]f_({\rvec}-{\rvec}_{\mu}(t_{n-1})),
\label{eqn:discreteG}
\end{eqnarray}
The integral in equation \ref{eqn:potential} becomes a sum, so the
potential can be computed as,
\begin{equation}
V_{\mu}({\rvec},t_{n}) := \sum_{\eta,\nu}\Delta x\Delta y e^{-|{\Delta x \eta\ivec+\Delta y \nu\jvec} - {\rvec}_{\mu}(t_{n-1})|/\sigma} G({\Delta x \eta\ivec+\Delta y \nu\jvec},t_{n}),
\label{eqn:discreteV}
\end{equation}
From the potential, the direction of the walker $\evec$ can be
computed. For convenience, we will write the numerator of
eqn. \ref{eqn:unitvector} as the un-normalized vector $\bar{\evec}$,
and then normalize it once it has been computed to calculate
$\evec$. The un-normalized $x$-component of eqn. \ref{eqn:unitvector}
becomes,
\begin{equation}
\bar{e}_{x\mu}(t_n) := \frac{(d_x-r_x)}{|{\dvec}-{\rvec}|} + \frac{[V_{\mu}(x_{\mu}+\Delta x,y_{\mu})  - V_{\mu}(x_{\mu}-\Delta x,y_{\mu})]}{2\Delta x},
\label{eqn:discreteEx}
\end{equation}
and the unnormalized $y$-component of eqn. \ref{eqn:unitvector} is
approximately,
\begin{equation}
\bar{e}_{y\mu}(t_n) := \frac{(d_y-r_y)}{|{\dvec}-{\rvec}|} + \frac{[V_{\mu}(x_{\mu},y_{\mu}+\Delta y)  - V_{\mu}(x_{\mu},y_{\mu}-\Delta y)]}{2\Delta y},
\label{eqn:discreteEy}
\end{equation}
Here $x_{\mu}$ and $y_{\mu}$ are defined via $\rvec_{\mu} =
x_{\mu}\ivec + y_{\mu}\jvec$. The components $\bar{e}_{x\mu}$ and
$\bar{e}_{y\mu}$ are combined into a single vector $\bar{\evec} =
\bar{e}_{x\mu}\ivec + \bar{e}_{y\mu}\jvec$ which is then normalized to
determine the direction of the next step,
\begin{equation}
{\evec}(t_n) := \bar{{\evec}}(t_n) / |\bar{{\evec}}(t_n)|
\label{eqn:discreteE}
\end{equation}
The average angle (equation \ref{eqn:averageangle}) is approximately,
\begin{equation}
\phi = \sum_{i=1}^{\Delta/\Delta t}\beta(t_{n-i}) \Delta t / \Delta
\end{equation}
with a similar scheme for the average of previous directions in
equation \ref{eqn:weightingnew}. Equation \ref{eqn:weightingnew} is
otherwise discretized in the same way as equation
\ref{eqn:discreteEx}.

The walker position is updated according to the discretized version of
eqn. \ref{eqn:velocity}.
\begin{equation}
{\rvec}_{\mu}(t_{n}) := {\rvec}_{\mu}(t_{n-1}) + \Delta t v_{0}{\evec}_{\mu}(t_{n}).
\label{eqn:discretePos}
\end{equation}
The vectors ${\rvec}_{\mu}$ were permitted to take any real value to
avoid truncation errors that could lead to limit cycles.

It is important to note that, in general, the validity of first order
integration schemes is not guaranteed. We checked our code and
approximate scheme against results in Refs. \refcite{helbing1997a} and
\refcite{helbing1997b}, finding no significant differences. We have
run simulations using 3 different resolutions of $\Delta x$, $\Delta
y$ and $\Delta t$ (see the results section of this article). The
scheme was found to converge quickly on reduction of the size of the
discrete steps.

As humans tend to walk with different step sizes, we varied the speed
of the individual walkers. This variation leads to continuous
paths. The width of the area studied was chosen as 10m and the length
of the observed area (in the direction of the incline) as 25m. This is
probably realistic for this type of trail formation, since we suspect
that walkers aim for local goals, rather than the final goal of the
peak (which is not necessarily visible from all locations).

\subsection{Algorithm one}

To clarify our computational scheme further, we detail the algorithms
that we use for computation. Two computational schemes have been
used. Algorithm one is based on Rules Ia and IIa, with algorithm two
based on Rules Ib and IIb. The steps in the algorithm are:

\begin{enumerate}
\item To initialize, set the ground condition to the default value
$G(\rvec,t_0)=0$.
\item Choose a walker speed following equation $v_{0\mu}=(0.5+r)$ms$^{-1}$, where $r$ is a random number between 0 and 1.
\label{step:selectvelocity}
\item Select the direction for the walker (i.e. up or down-hill)
randomly with equal probability.
\item Initialize the walker position at the bottom or top of the
incline depending on whether an up or down walker has been selected.
\item Compute the attractive potential $V$ from equation
\ref{eqn:discreteV}.
\label{step:computeV}
\item Determine the default walker direction from equation \ref{eqn:discreteE}.
\item Apply the persistence of direction formula, equation \ref{eqn:weighting}
\item Apply the forbidden angle rule. If the angle, $\gamma$, lies
within a forbidden zone ($|\gamma|<\theta$), then the vector $\evec$
is mapped to the closest angle outside the forbidden region. In the
event that $\gamma=0$ our walker has a preference to move left.
\item Update the walker position using equation \ref{eqn:discretePos}.
\item Update the ground condition using equation \ref{eqn:discreteG}.
\item Repeat from step \ref{step:computeV} until the walker has reached the destination.
\item Repeat from step \ref{step:selectvelocity} until the action of
sufficient walkers on the ground condition has been investigated.
\end{enumerate}

\subsection{Algorithm two}

\begin{enumerate}
\item Initialize the ground condition.
\item Select the speed of the walker.
\label{step:selectvelocitytwo}
\item Select the direction of the walker.
\item Initialize the walker position.
\item Compute $V$ from equation \ref{eqn:discreteV}.
\label{step:computeVtwo}
\item Apply the alternative formula for persistence of direction, equation
\ref{eqn:weightingnew}.
\item Apply the forbidden angle rule. If the angle, $\beta$, lies
within a forbidden zone ($|\beta|<\theta$), then the vector $\evec$
is mapped to the closest angle outside the forbidden region. In the
event that $\gamma=0$ our walker has a preference to move left.
\item Update the walker position $\rvec_{\mu}$ using equation \ref{eqn:discretePos}.
\item Update $G$ using equation \ref{eqn:discreteG}.
\item Repeat from step \ref{step:computeVtwo} until the destination has been reached.
\item Repeat from step \ref{step:selectvelocitytwo} until the desired
number of walkers have traversed the slope.
\end{enumerate}

The simulation was implemented in c++. The random number generator,
 ran2 from `Numerical Recipes in C' was used to ensure a long period
 \cite{press1992a}. We tested our code against the triangular and
 square examples in Ref. \refcite{helbing1997b}.

\section{Results}
\label{sec:results}

\subsection{Algorithm one}

In this section, results from the mountain walker extension to the
active walker model are shown. We begin by considering only walkers
traveling down the slope. The following parameters were used for the
simulations: the maximum ground potential was set to $G_{\rm
max}=200$m$^{-1}$ (the units of $G$ are set by Eq.
\ref{eqn:potential} and Eq. \ref{eqn:unitvector}) while the minimum
ground potential was set to $G_0=0$. Larger values of $G_{\rm max}$
lead to larger attraction to the path. A new walker descended the
inclined area as soon as the existing walker reached the destination
(tests showed that there was little difference between this scheme and
starting a walker every 100s). The visibility, $\sigma$ of the paths
was set to 10m and the intensity, $I$ was set to $l^2 G_{\rm max} / N$,
where $N$ is the number of footprints needed to wear the ground
condition to $1/e$ of its maximum value. $N$ was set to 50 footfalls
and the weathering parameter was initially set to $T=1000$s. The
values of $N$ and $T$ are smaller than in a real trail system, where
$N$ would be of the order of several hundred footfalls and $T$ would
be of the order of a few days. This still leads to realistic
simulations since Helbing {\it et al.} have found that a combination
of several parameters of the active walker model could be represented
by the single parameter $\kappa=IT/\sigma=G_{\max}T/N\sigma$
\cite{helbing1997a}. Individual walkers were assigned a random speed
between 0.5m/s and 1.5m/s.  In the simulation, walkers were given the
starting position ${\rvec}_{\rm initial}=(0{\rm m}, 5{\rm m})$ and a
destination of ${\rvec}_{\rm final} = (25{\rm m}, 5{\rm m})$ with the
$x$-direction being the distance down the incline and $y$-direction
the distance across the incline. Initially, 25000 walkers traversed the incline
in each simulation.

In our initial simulations, we set the parameter $\alpha = 0$,
consistent with the standard setup of the active walker model. This
led to a curious result, as shown in Fig. \ref{fig:mountain1}. We
found that walkers changed their walking direction frequently, with
some walkers changing direction on every time step. This waddling gait is
not very comfortable on an incline, and it is likely to make the
walker very unstable. Also, the speed of advance down a hill is likely
to be quite slow. We thoroughly investigated the parameter space to
look for the zigzag paths, but did not find them. As this waddling
gait is not a practical way to descend a mountain, it is reasonable to
conclude that the active walker model is missing some crucial physics:
a walker is much more likely to ambulate by moving consecutive legs in
similar directions, otherwise the resulting motion will be very
unstable. For this reason, we introduce a factor, $\alpha$, to
represent persistence of direction (see Eq. \ref{eqn:weighting}).
$\alpha$ controls a weighted average between the angles on consecutive
time steps.

By changing the weighting factor, which alters the influence of the
recent direction of motion on the walker, and by changing the size of
the forbidden angle $\theta$, we were able to estimate how large
$\alpha$ needs to be to produce convincing zigzag paths. Again, we
assumed that walkers always traveled down the incline (in the
direction of increasing $x$) and otherwise used the same
parameters. We ran simulations with $\theta$ ranging from $\theta=5^o$
to $\theta=45^o$ and $\alpha$ ranging from $\alpha = 0.1$ to $\alpha =
0.9$.

We show the results of single runs for a minimum safe angle of $\theta
= 5^o$ in figure \ref{fig:mountainwalkerfive}. The zig-zag paths are
immediately apparent. Multiple turns in the path can be seen when the
parameter, $\alpha\gtrsim 0.5$
(Fig. \ref{fig:mountainwalkerfive}(b)). This indicates that when
walkers choose where to place the next footfall, around half of the
choice comes from a desire to maintain the previous walking direction,
and half from the wish to reach the destination quickly via a
path. The effect of reducing the influence of the previous angle can
be seen in Fig. \ref{fig:mountainwalkerfive}(a) where $\alpha =
0.4$. As a result of the reduced influence of the previous direction,
the walker's motion follows only a slight wiggle. In panel (c) of
Fig. \ref{fig:mountainwalkerfive}, where $\alpha = 0.6$ and the
influence of the previous angle is increased, the overall size of
features increases. This is a common feature of all simulations where the influence of previous step directions is increased through $\alpha$.

As the minimum safe angle is increased to $15^o$, we note that the
minimum values of $\alpha$ which are needed to generate zig-zag
patterns decreases slightly. In
Fig. \ref{fig:mountainwalkerfifteen}(a) where $\alpha = 0.4$, we
observe that a single zigzag emerges in the path. As walkers get
closer to their destination, the rate at which they change their
walking direction increases. Fig. \ref{fig:mountainwalkerfifteen}(b)
shows a similar path for $\alpha = 0.45$. In
Fig. \ref{fig:mountainwalkerfifteen}(c) we observe that when $\alpha =
0.5$ walkers take larger detours away from the direct path between
entry and exit, with several changes of direction along the
trail. Similar results were found when $\theta=25^{o}$ (Fig.
\ref{fig:mountainwalkertwentyfive}). While zigzag forms develop during
the simulations, the rather diffuse paths are not satisfactory
representations of mountain trails.

\begin{figure*}
\includegraphics[height=63mm,angle=270]{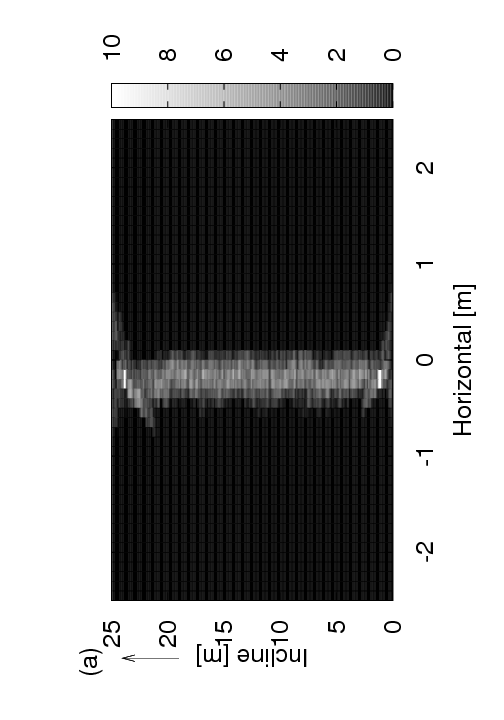}
\includegraphics[height=63mm,angle=270]{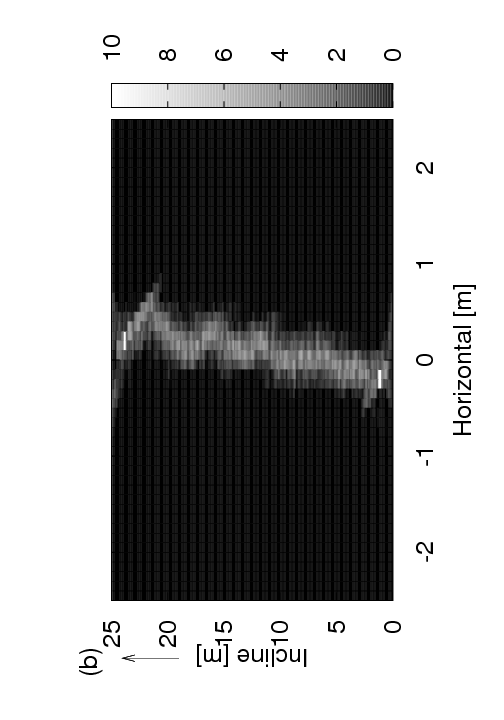}
\includegraphics[height=63mm,angle=270]{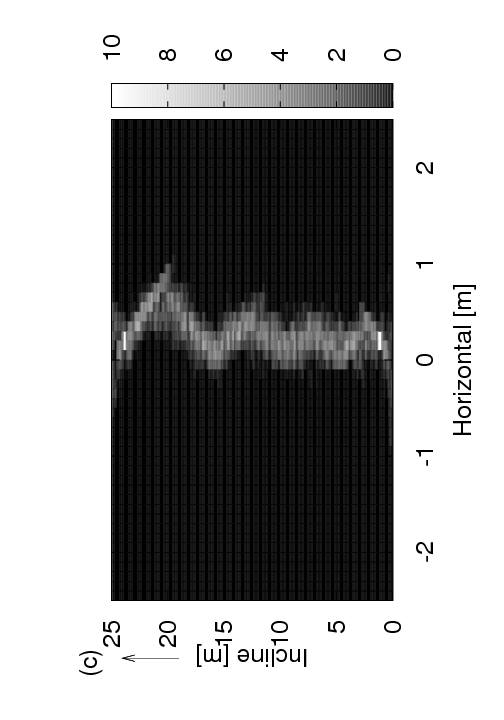}
\includegraphics[height=63mm,angle=270]{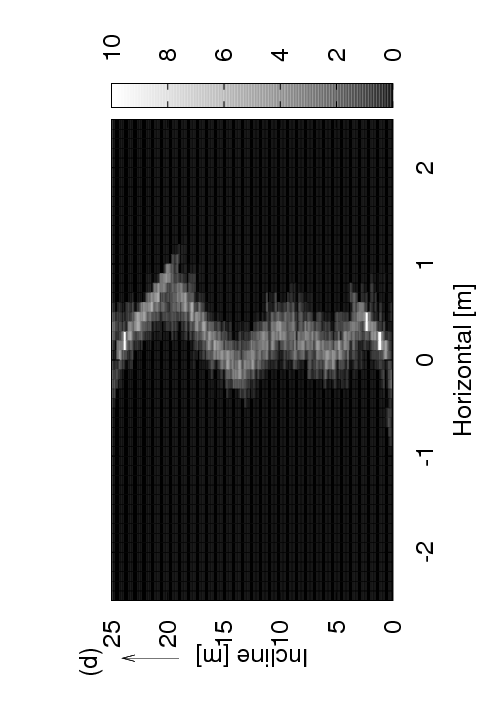}
\includegraphics[height=63mm,angle=270]{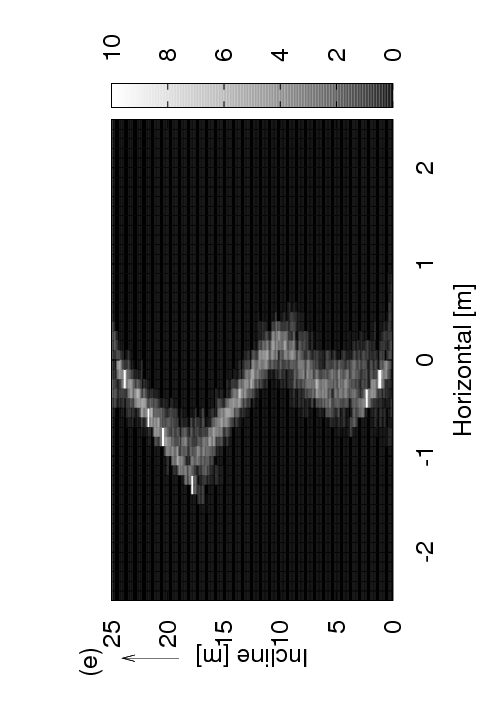}
\includegraphics[height=63mm,angle=270]{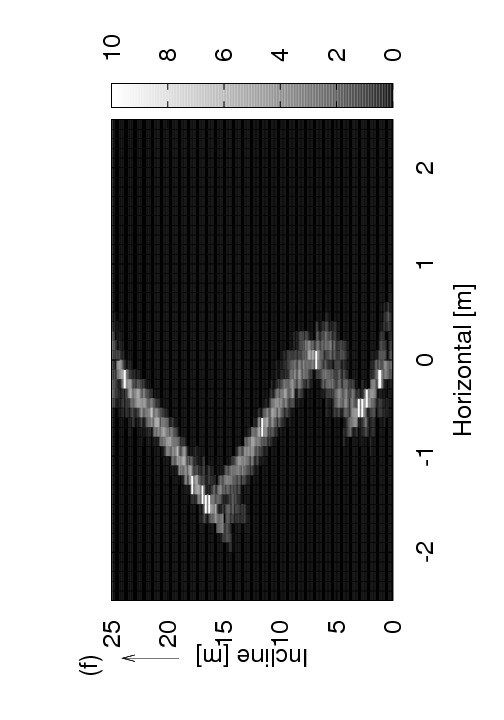}
\includegraphics[height=63mm,angle=270]{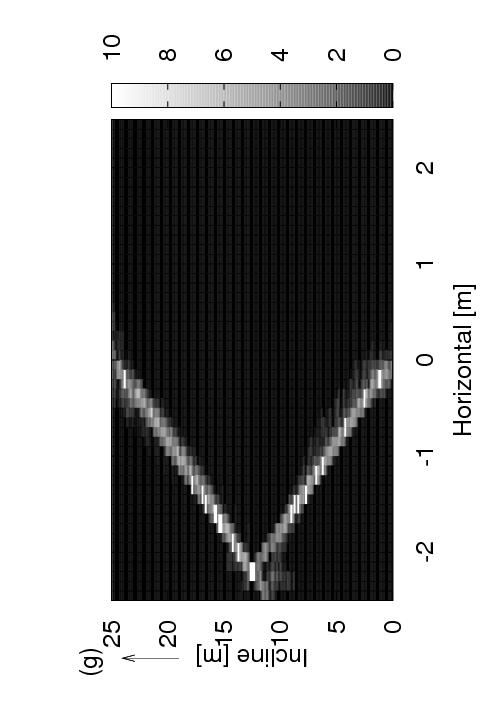}
\includegraphics[height=63mm,angle=270]{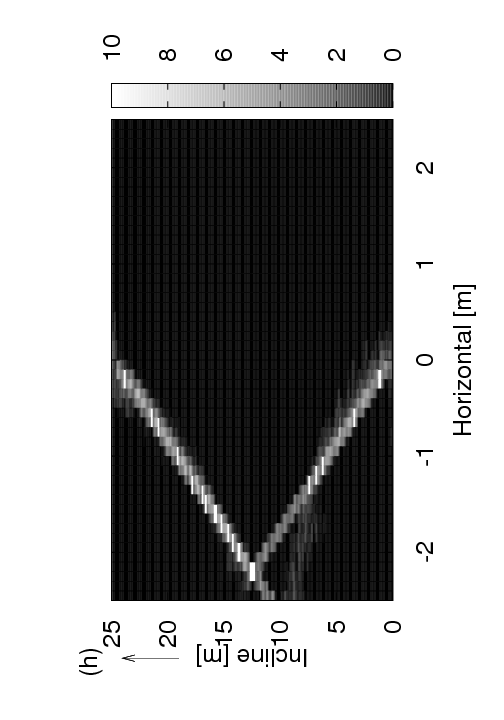}
\caption{The results of simulations of 2500 walkers traveling consecutively up and down the incline. Algorithm one is used (a) $\alpha=0.30$ (b) $\alpha=0.35$ (c) $\alpha=0.40$ (d) $\alpha=0.45$ (e) $\alpha=0.50$ (f) $\alpha=0.55$ (g) $\alpha=0.60$ (h) $\alpha=0.65$. $T=1500$s, $N=50$, $\sigma=10$m and $v=(0.5+r)$ms$^{-1}$ where $r\in[0,1)$ is a random variate. We also considered simulations with 25000 walkers, with no noticeable difference to the pattern of paths formed. $\Delta t=1.0$s}
\label{fig:mountainwalkertwoway}
\end{figure*}


\begin{figure*}
\includegraphics[height=63mm,angle=270]{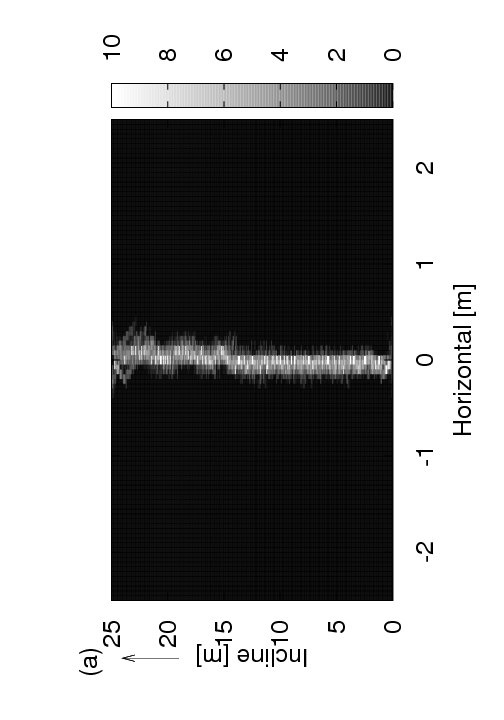}
\includegraphics[height=63mm,angle=270]{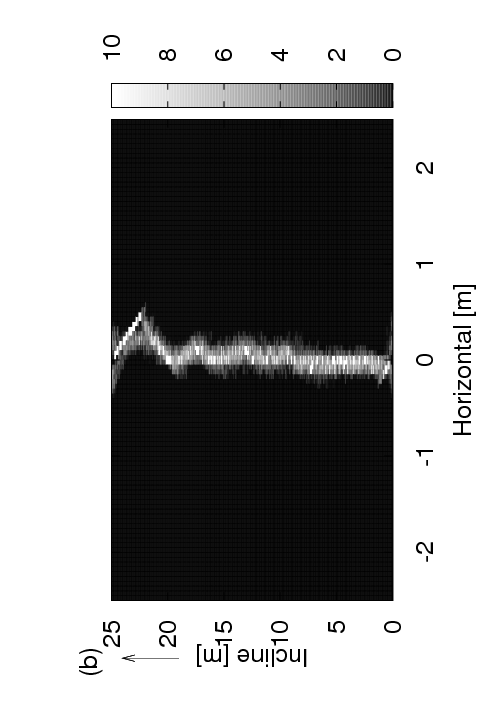}
\includegraphics[height=63mm,angle=270]{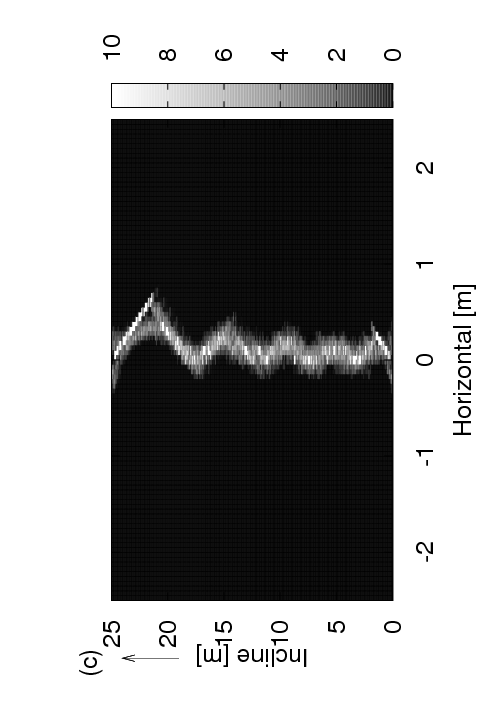}
\includegraphics[height=63mm,angle=270]{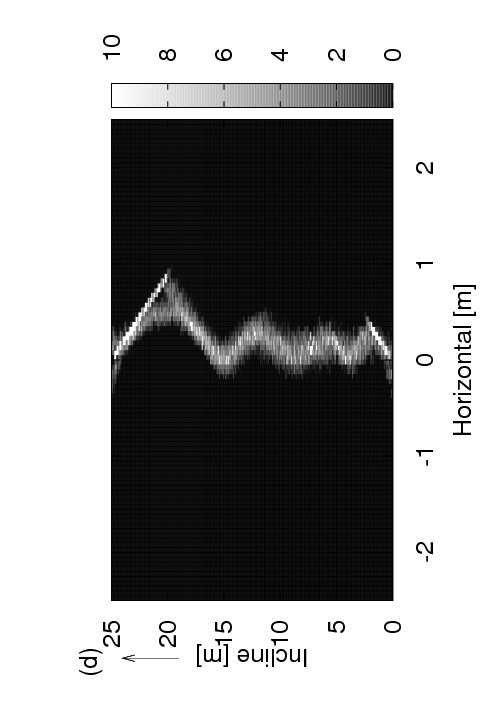}
\includegraphics[height=63mm,angle=270]{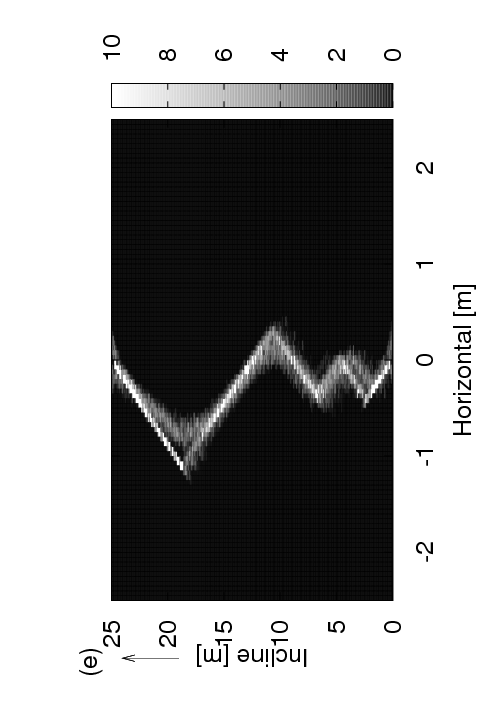}
\includegraphics[height=63mm,angle=270]{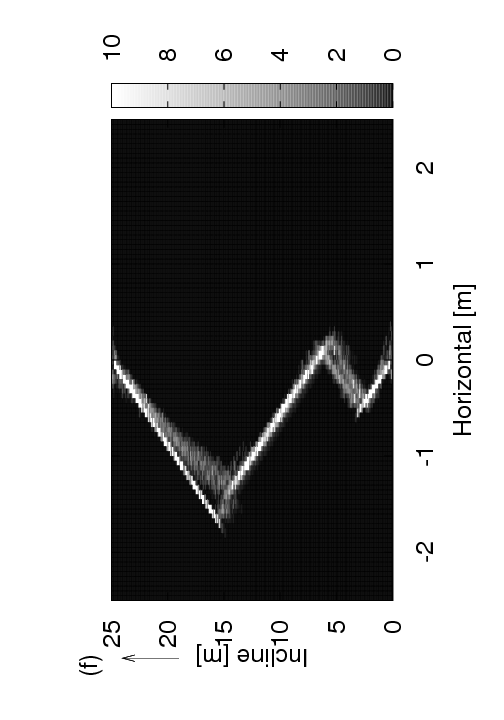}
\includegraphics[height=63mm,angle=270]{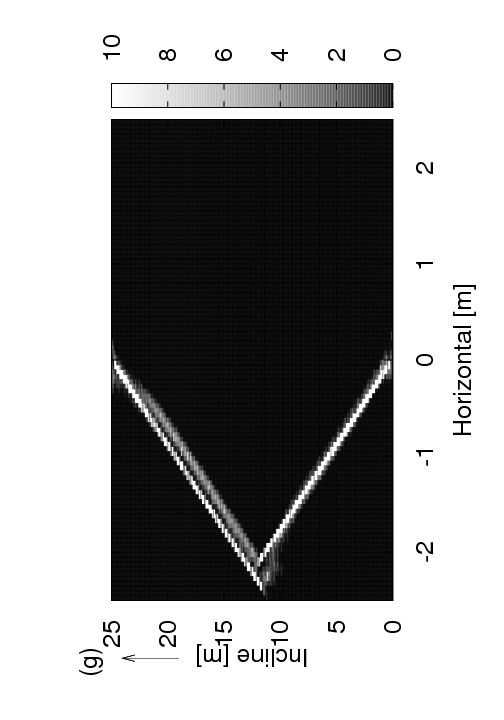}
\includegraphics[height=63mm,angle=270]{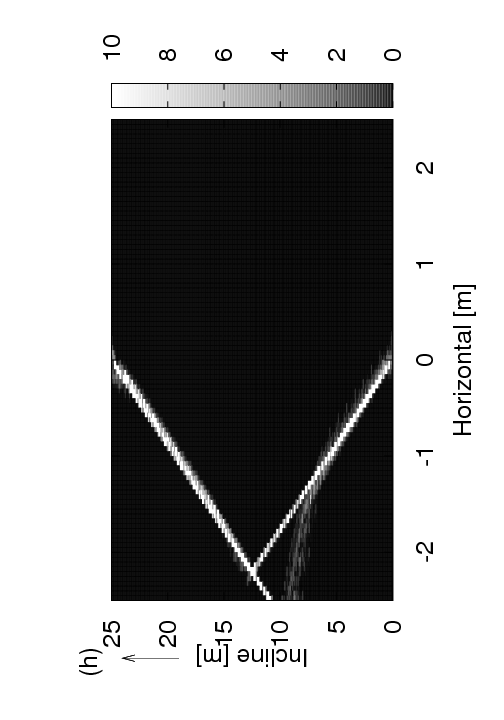}
\caption{As Fig. \ref{fig:mountainwalkertwoway} with double the space
and time resolution used in the computation $\Delta x=\Delta y = 5$cm
and $\Delta t=0.5$s. The shapes of the paths are robust against the
finite size scaling.}
\label{fig:mountainwalkertwowayhalfsec}
\end{figure*}

\begin{figure*}
\includegraphics[height=63mm,angle=270]{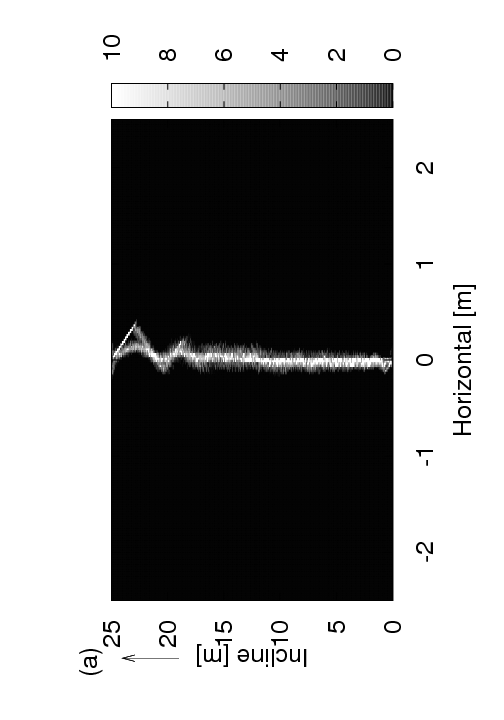}
\includegraphics[height=63mm,angle=270]{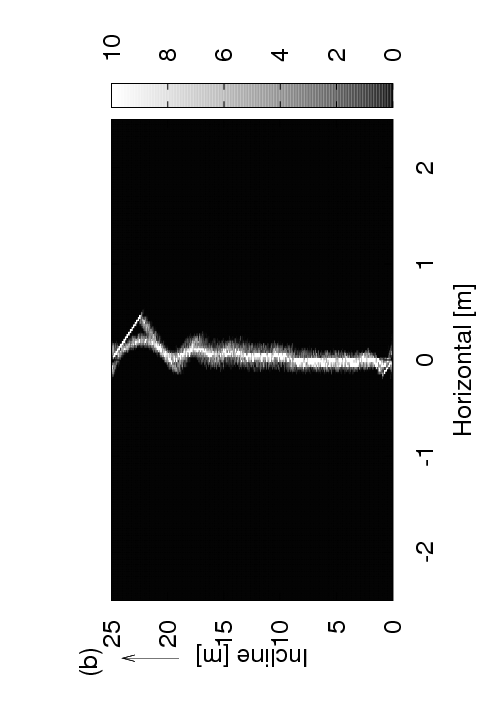}
\includegraphics[height=63mm,angle=270]{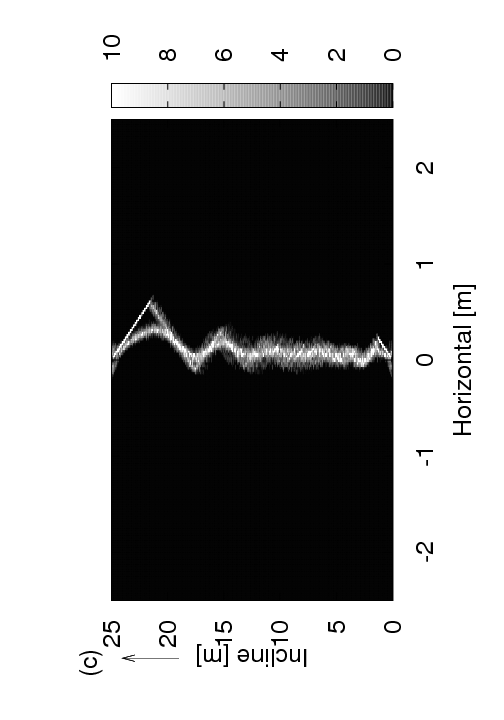}
\includegraphics[height=63mm,angle=270]{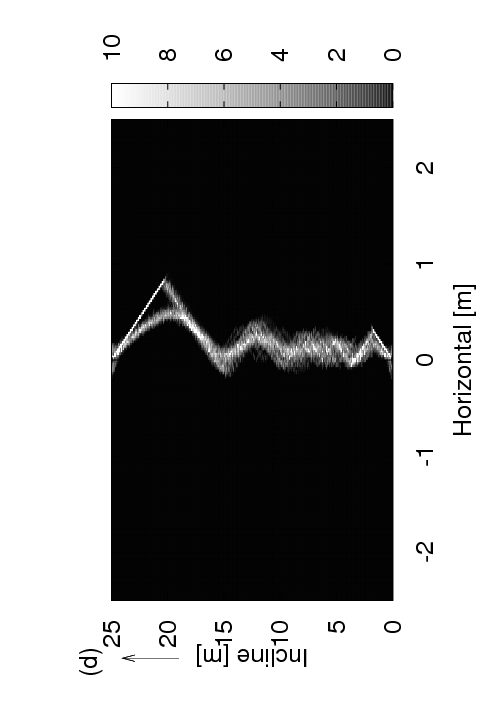}
\includegraphics[height=63mm,angle=270]{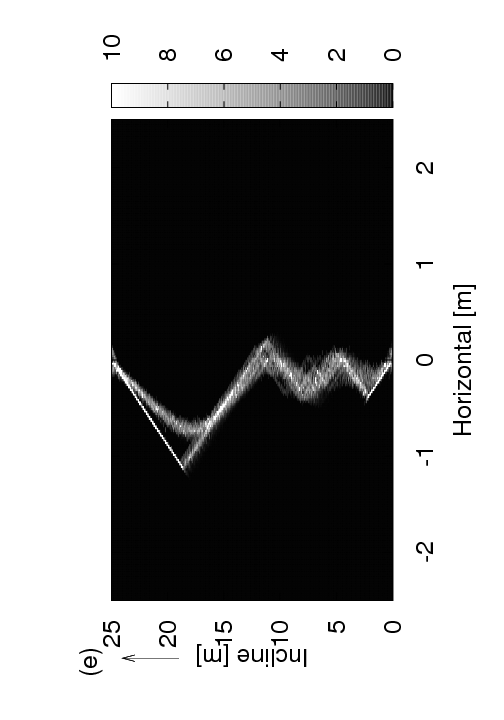}
\includegraphics[height=63mm,angle=270]{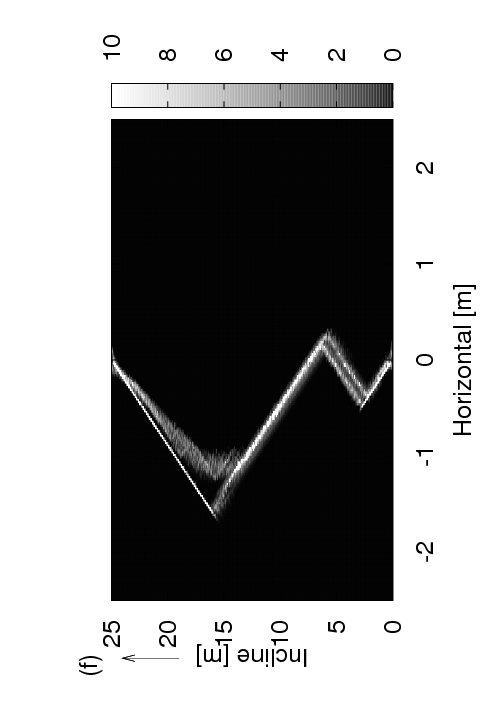}
\includegraphics[height=63mm,angle=270]{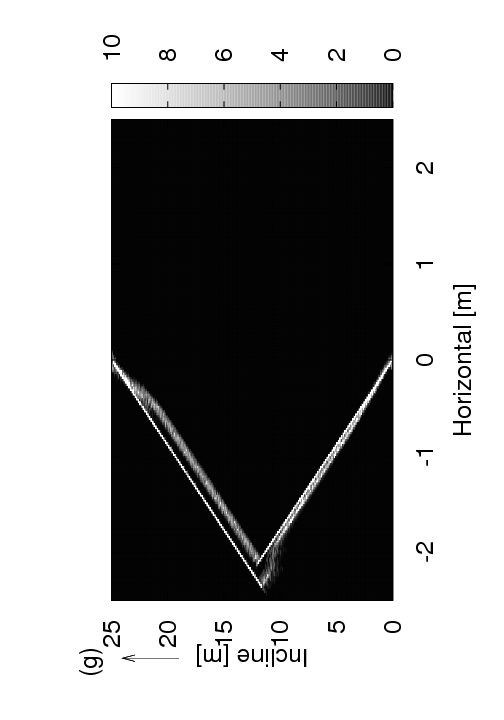}
\includegraphics[height=63mm,angle=270]{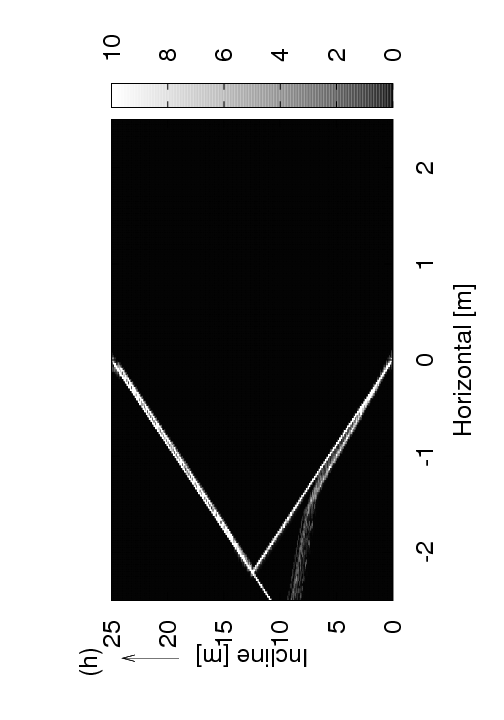}
\caption{As Fig. \ref{fig:mountainwalkertwoway} with four times the
space and time resolution used in the computation $\Delta x=\Delta y =
2.5$cm and $\Delta t=0.25$s. Again, the shapes of the paths are robust
against the finite size scaling.}
\label{fig:mountainwalkertwowayquartsec}
\end{figure*}


\begin{figure*}
\includegraphics[height=63mm,angle=270]{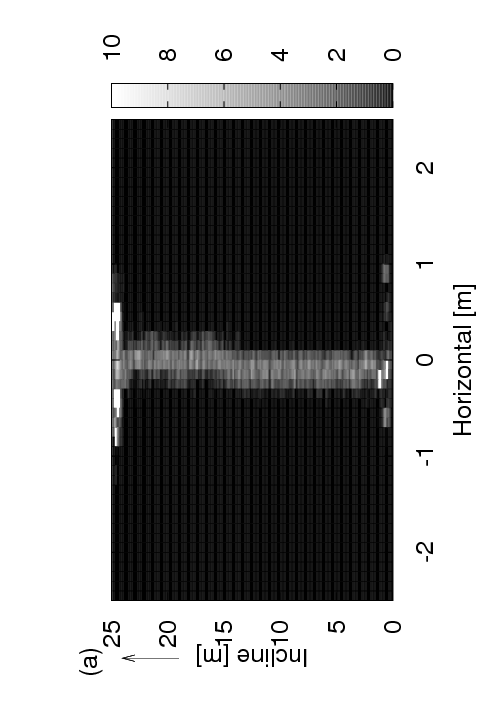}
\includegraphics[height=63mm,angle=270]{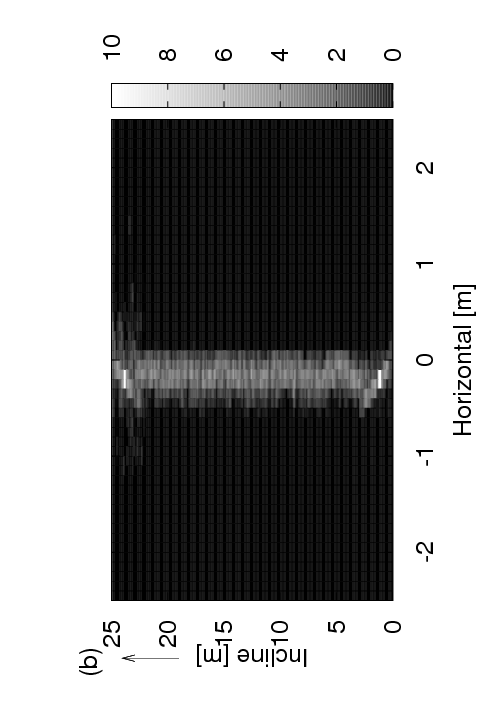}
\includegraphics[height=63mm,angle=270]{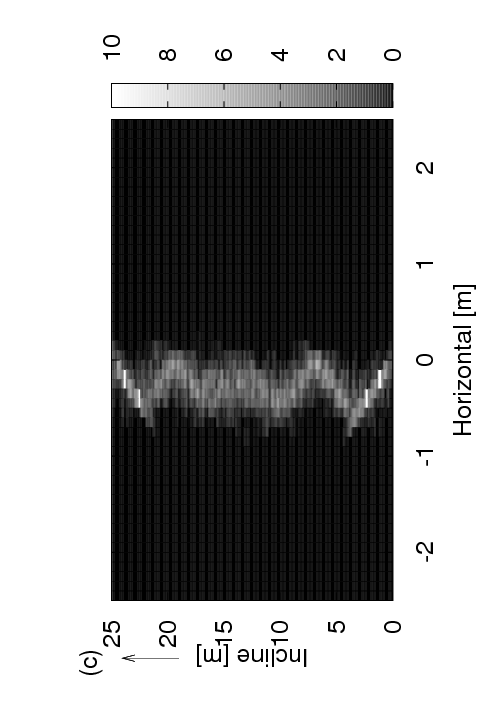}
\includegraphics[height=63mm,angle=270]{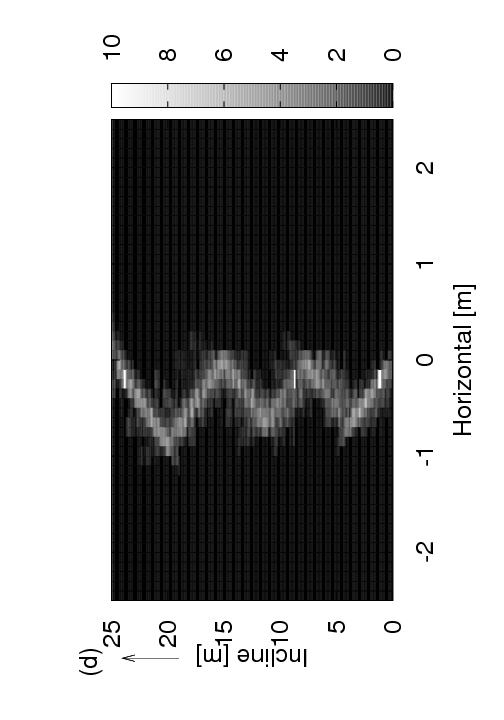}
\includegraphics[height=63mm,angle=270]{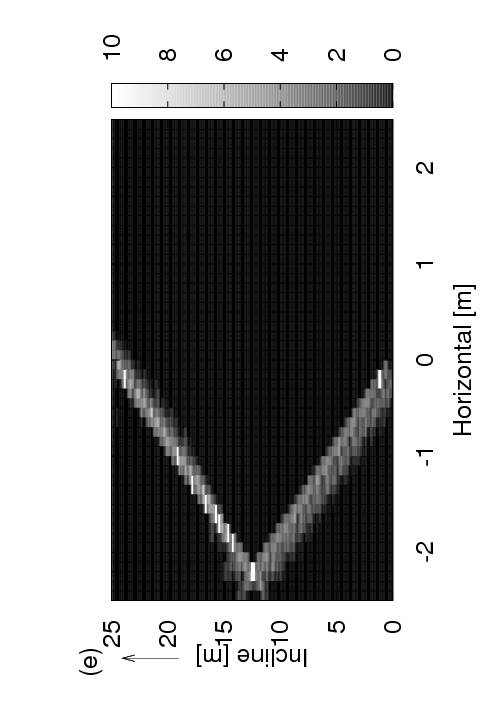}
\includegraphics[height=63mm,angle=270]{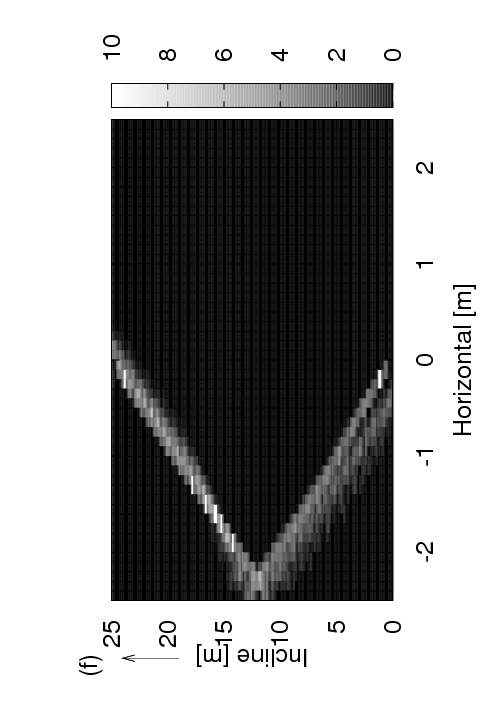}
\caption{The alternative direction averaging scheme for the mountain
walker. (a) $\bar{\alpha}=0.5$, (b) $\bar{\alpha}=1.0$, (c)
$\bar{\alpha}=1.5$, (d) $\bar{\alpha}=2.0$, (e) $\bar{\alpha}=2.5$,
(f) $\bar{\alpha}=3.0$. $\Delta t=1$s, $\Delta x = \Delta y =
10$cm. Results from both schemes are in qualitative agreement. The
size of features increases monotonically with $\bar{\alpha}$.}
\label{fig:mountainwalkertwowaynewscheme}
\end{figure*}


\begin{figure*}
\includegraphics[height=63mm,angle=270]{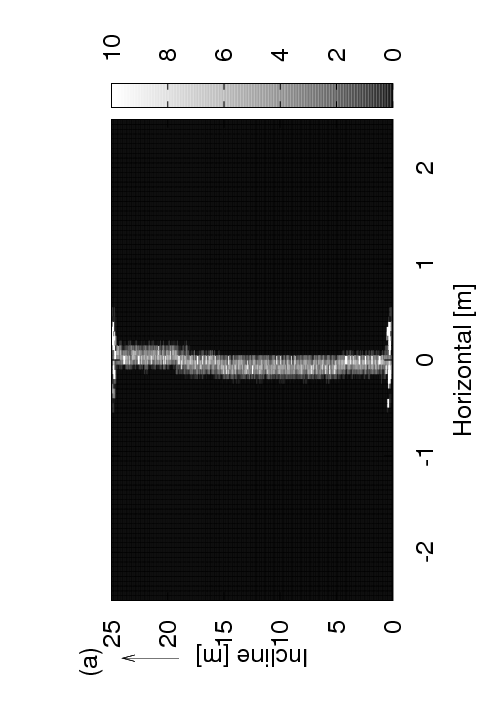}
\includegraphics[height=63mm,angle=270]{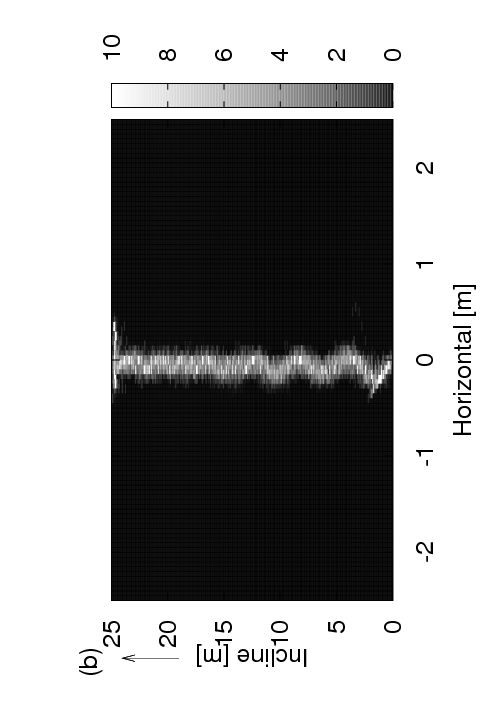}
\includegraphics[height=63mm,angle=270]{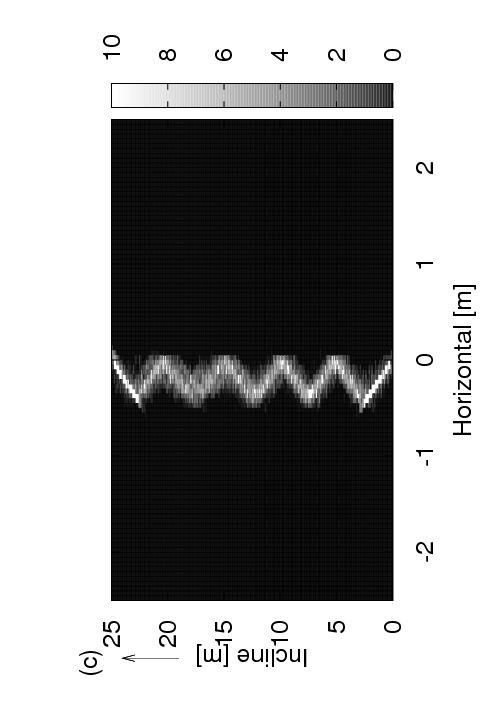}
\includegraphics[height=63mm,angle=270]{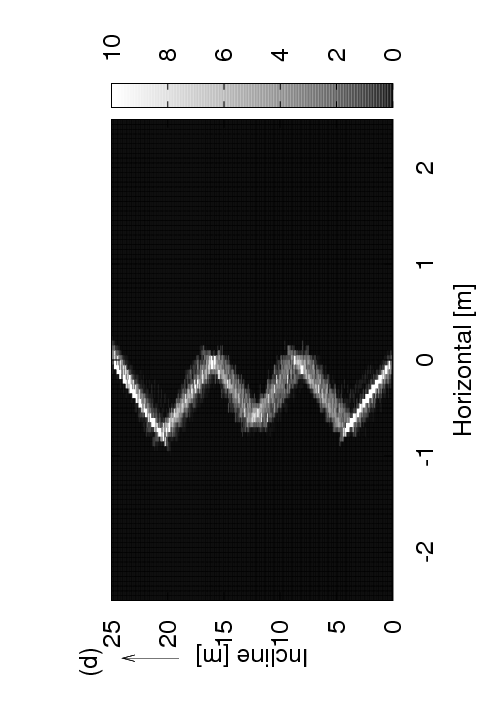}
\includegraphics[height=63mm,angle=270]{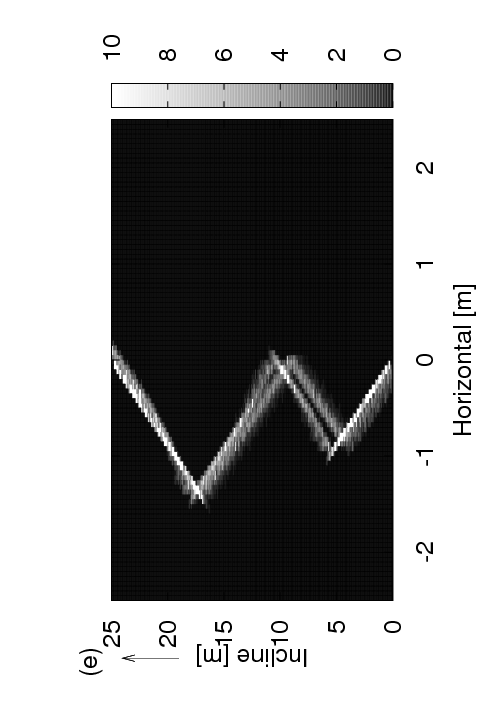}
\includegraphics[height=63mm,angle=270]{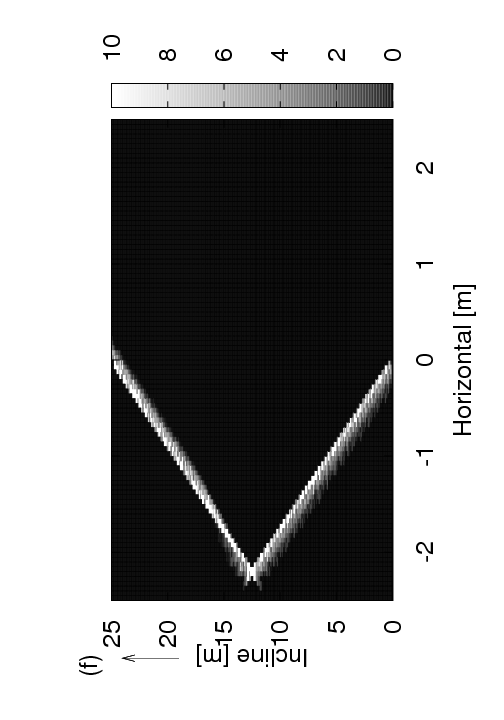}
\caption{As figure \ref{fig:mountainwalkertwowaynewscheme}, except $\Delta t=0.5$s, $\Delta x = \Delta y = 5$cm.}
\label{fig:mountainwalkertwowaynewschemehalfsec}
\end{figure*}

\begin{figure*}
\includegraphics[height=63mm,angle=270]{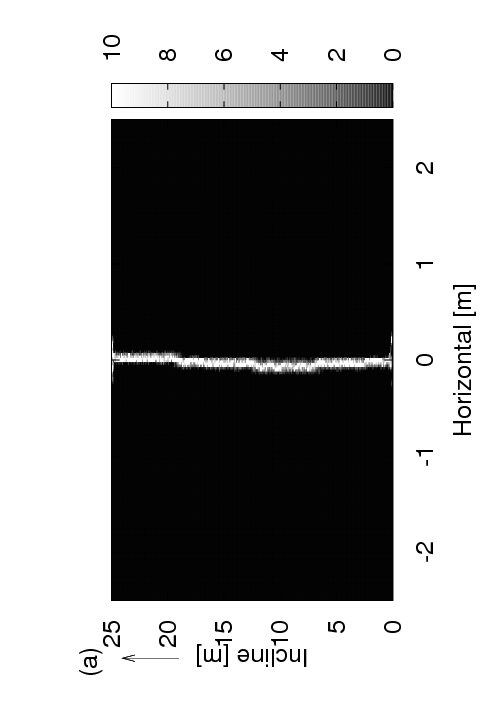}
\includegraphics[height=63mm,angle=270]{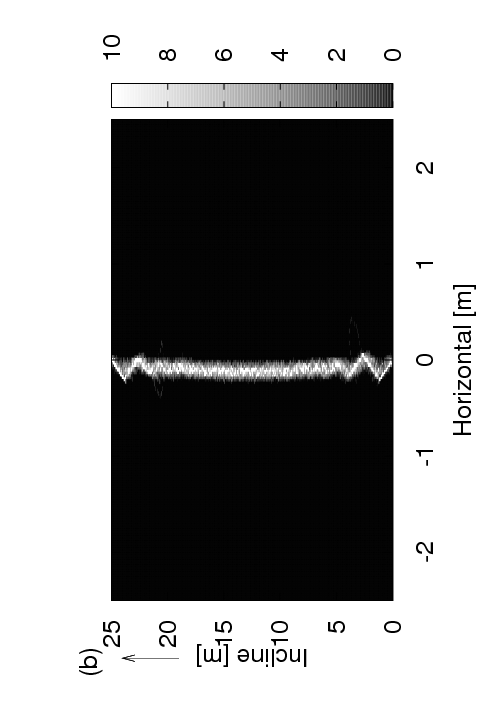}
\includegraphics[height=63mm,angle=270]{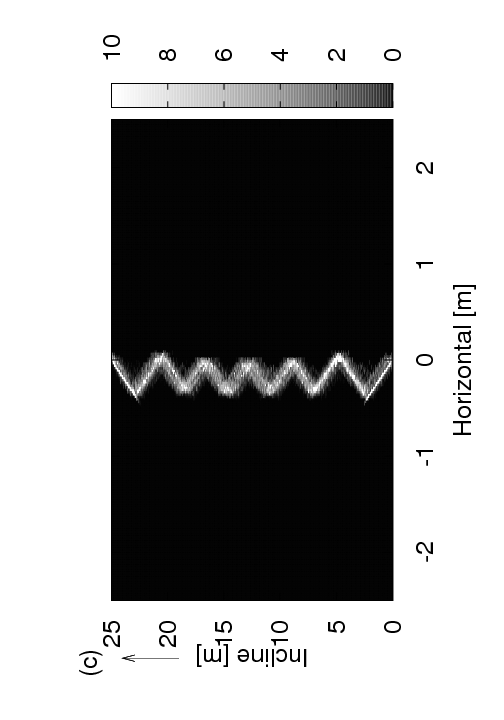}
\includegraphics[height=63mm,angle=270]{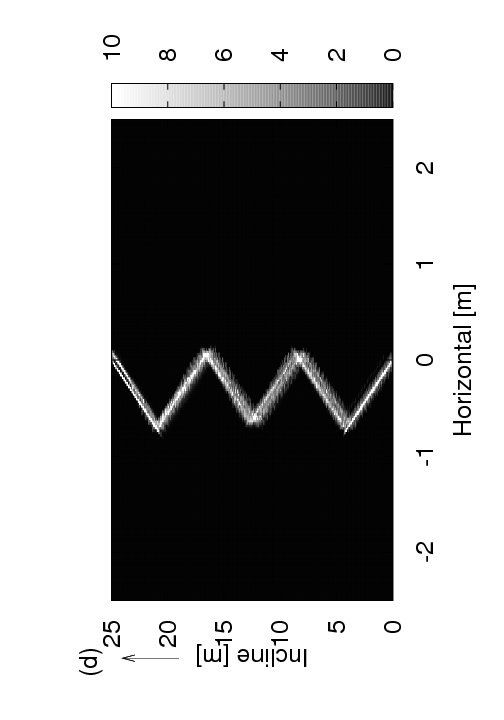}
\includegraphics[height=63mm,angle=270]{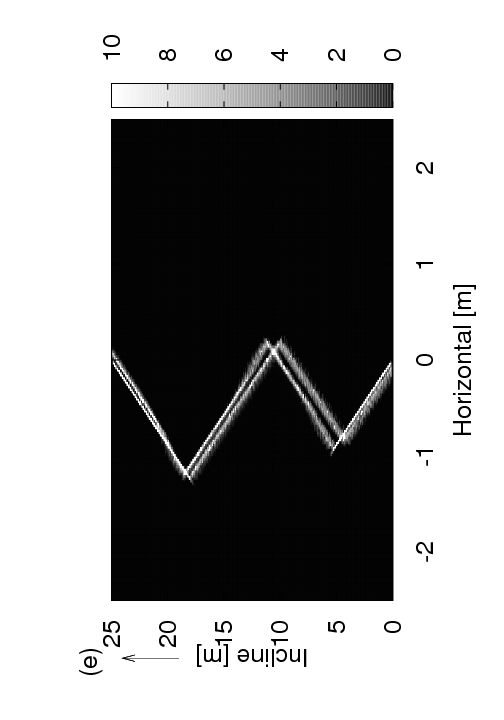}
\includegraphics[height=63mm,angle=270]{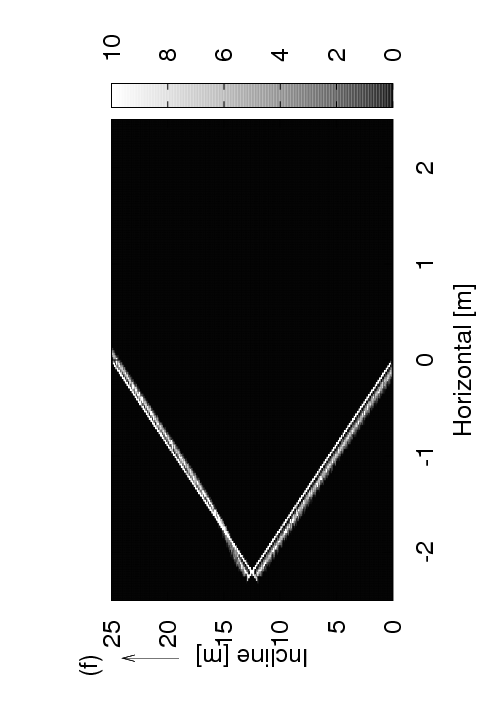}
\caption{As figure \ref{fig:mountainwalkertwowaynewscheme}, except $\Delta t=0.25$s, $\Delta x = \Delta y = 2.5$cm.}
\label{fig:mountainwalkertwowaynewschemequartsec}
\end{figure*}

We complete our simulations using algorithm one, by examining walkers
that travel consecutively up and down the incline. Results of the
simulations can be seen in figure \ref{fig:mountainwalkertwoway}. A
slightly different weathering time of $T=1500$s was used. There is no
requirement that the forbidden angles are the same for hikers moving
up and down the incline. In general, we expect that the different
mechanisms for walking up and down inclines lead to different
forbidden angles. In this set of simulations, we choose a forbidden
angle of $\theta_1=25^{o}$ for walkers moving up the diagram (down the
gradient). Those moving down the diagram (up the gradient) have a
minimum safe angle of $\theta_2=10^o$. We run simulations for a range
of $\alpha$. Some zig-zag patterns can be seen for low alpha, but
patterns of decent size do not appear until $\alpha \gtrsim 0.45$
(panel d), Again, the sizes of the bends in the path increase in size
with $\alpha$. The combined effect of walkers moving in both
directions are trails with zig-zag patterns that have a similar angle
to the largest of the two forbidden angles. The walkers with a smaller
forbidden angle round off the sharp turns in the paths that were found
on the zig-zags formed when walkers are only permitted to move in a
single direction. This rounding is a direct consequence of the
attraction term in the active walker model and may explain the curved
nature of spontaneously formed mountain trails. Moreover, the
inclusion of walkers with different minimum safe angles leads to well
defined trails (rather than the diffuse trails found previously). This
is probably because the walkers with smaller forbidden angles have
more freedom to change direction, allowing the active walker rules to
function effectively.

To test the effects of discretization, we compute results using
smaller $\Delta t=0.5$s, $\Delta x=5$cm and $\Delta y=5$cm in figure
\ref{fig:mountainwalkertwowayhalfsec}. In figure
\ref{fig:mountainwalkertwowayquartsec} we use even smaller steps
$\Delta t=0.25$s, $\Delta x=2.5$cm and $\Delta y=2.5$cm. The absence
of new features shows our calculations to be stable against finite
size scaling.

\subsection{Algorithm two}

We also run simulations using algorithm two. The results are shown in
Figs. \ref{fig:mountainwalkertwowaynewscheme},
\ref{fig:mountainwalkertwowaynewschemehalfsec} and
\ref{fig:mountainwalkertwowaynewschemequartsec} with the resolution of
the discretization increased between figures.  2500 walkers traversed
the incline in both directions. The parameters of the runs were
$T=1500$s, $N=50$, $\sigma=10$m and $v_{0\mu}=(0.5+r)$ms$^{-1}$ where
$r\in[0,1)$ is a random variate. $\bar{\alpha}$ was varied. Again,
we choose $\theta_1=25^{o}$ and $\theta_2=10^o$. The results from both
algorithms one and two are in good qualitative agreement. Note that
the results for $\bar{\alpha}=2.5$ required slightly higher resolution
for convergence. In comparison to the patterns from algorithm one, the
zigzags are quite regular. Small wiggles of around $0.5$metres across
form when $\bar{\alpha}=1.5$. As before, the features become larger as
$\bar{\alpha}$ is increased.

\section{Summary}
\label{sec:summary}

We have developed an extension to the active walker model to handle
the formation of trails on inclines. Our simulations are in
qualitative agreement with empirical observations of mountain path
formations. Such trails are characterized by a zig-zag pattern. Our
extension took account of the inability of walkers to walk directly up
or down very steep gradients. We have shown that some supplementary
rules need to be included in the active walker model to achieve
features consistent with mountain trails. Those additional rules are
that (a) the consecutive steps of walkers tend to be in the same
direction and that (b) there is a maximum permitted angle of motion
down an incline to avoid falling and (c) there is a maximum angle of
ascent for physiological reasons such as limited ankle
flexibility. When rules to encourage consecutive steps are not
present, we find that walkers travel in an unusual manner, only taking
a single step before changing direction. We also found that the
presence of walkers moving both uphill and downhill (with different
forbidden angles) is important for forming well defined zig-zag
paths. Walkers traveling downhill with a larger forbidden angle are
constrained to form zigzags but create rather diffuse paths unless
those paths are complemented by walkers traveling uphill. This happens
because the walkers with more angular freedom become attracted to and
reinforce the paths.


The preference of walkers to take consecutive steps in the same
direction could also be relevant for the understanding of trail
formation in flat areas, and it is our opinion that a persistence of
direction parameter should also be included when simulating trail
systems on level ground. Thus, our main conclusions are that (1)
correlation between consecutive step directions should be included in
the active walker model and that (2) an extension of the active walker
model to include forbidden angles, correlation between consecutive
steps and the combination of walkers moving both up- and down-hill is
suitable to understand the formation of mountain paths.
 
\section*{Acknowledgments}

We are pleased to acknowledge Feo Kusmartsev for useful
discussions. JPH would also like to thank Chloe Long for assistance
with photography.

\bibliographystyle{plain}
\bibliography{mountainwalker_v2_0}

\begin{thebibliography}{10}

\bibitem{alexander2002a}
R.~M. Alexander.
\newblock {\em Am. J. Hum. Biol.}, 14:641, 2002.

\bibitem{bell1973a}
K.~L. Bell and L.~C. Bliss.
\newblock {\em Biological Conservation}, 5:25, 1973.

\bibitem{coleman1981a}
R.~Coleman.
\newblock {\em Applied Geography}, 1:121, 1981.

\bibitem{goldstone2006a}
R.~L. Goldstone, A.~Jones, and M.~E. Roberts.
\newblock {\em IEEE Transactions on Systems, Man and Cybernetics, Part A},
  36:611, 2006.

\bibitem{goldstone2006b}
R.~L. Goldstone and M.~E. Roberts.
\newblock {\em Complexity}, 11:43, 2006.

\bibitem{helbing2001b}
D.~Helbing.
\newblock {\em Rev. Mod. Phys.}, 73:1067, 2001.

\bibitem{helbing1997a}
D.~Helbing, J.~Keltsch, and P.~Molnar.
\newblock {\em Nature}, 388:47, 1997.

\bibitem{helbing2001a}
D.~Helbing, P.~Molnar, I.~Farkes, and K.~Bolay.
\newblock {\em Environment and Planning B: Planning and Design}, 28:361, 2001.

\bibitem{helbing1997b}
D.~Helbing, F.~Schweitzer, J.~Kelysch, and P.~Molnar.
\newblock {\em Phys. Rev. E}, 56:2527, 1997.

\bibitem{kirchner2004a}
A.~Kirchner, H.~Klupfel, K.~Nishinari, A.~Schadschneider, and M.~Schreckenberg.
\newblock {\em J. Stat. Mech}, 2004:P10011, 2004.

\bibitem{leroux2002a}
A.~Leroux, J.~Fung, and H.~Barbeau.
\newblock {\em Gait and Posture}, 15:6474, 2002.

\bibitem{mcintosh2006a}
A.~S. McIntosh, K.~T. Beatty, L.~N. Dwan, and D.~R. Vickers.
\newblock {\em Journal of Biomechanics}, 39:2491, 2006.

\bibitem{press1992a}
W.~H. Press, S.~A. Teukolsky, W.~T. Vetterling, and B.~P. Flannerty.
\newblock {\em Numerical Recipes in C: The art of Scientific Computing}.
\newblock Cambridge University Press, 1992.

\bibitem{watts1992a}
A.~Watts.
\newblock {\em Climber's guide to Smith rock}.
\newblock Falcon Guides, Guilford, CT, USA, 1992.

\bibitem{whittle2001a}
M.~W. Whittle.
\newblock {\em Gait analysis: An introduction}.
\newblock Butterworth-Heinemann Ltd., 2001.

\bibitem{zheng1990a}
Y.~F. Zheng and J.~Shen.
\newblock {\em IEEE Transactions on Robotics and Automation}, 6:86, 1990.

\end{thebibliography}

\end{document}